# THE 37-MONTH MAXI/GSC SOURCE CATALOG IN THE HIGH GALACTIC-LATITUDE SKY

Kazuo Hiroi,[1] Yoshihiro Ueda,[1] Masaaki Hayashida,[1] Megumi Shidatsu,[1] Ryosuke Sato,[1] Taiki Kawamuro,[1] Mutsumi Sugizaki,[2] Satoshi Nakahira,[3] Motoko Serino,[2] Nobuyuki Kawai,[4] Masaru Matsuoka,[2,3] Tatehiro Mihara,[2] Mikio Morii,[4] Motoki Nakajima,[5] Hitoshi Negoro,[6] Takanori Sakamoto,[7] Hiroshi Tomida,[3] Yohko Tsuboi,[8] Hiroshi Tsunemi,[9] Shiro Ueno,[3] Kazutaka Yamaoka,[10] Atsumasa Yoshida,[7] Masato Asada,[6] Satoshi Eguchi,[11] Takanori Hanayama,[12] Masaya Higa,[8] Kazuto Ishikawa,[4] Masaki Ishikawa,[13] Naoki Isobe,[14] Mitsuhiro Kohama,[15] Masashi Kimura,[3] Kumiko Morihana,[2] Yujin E. Nakagawa,[14] Yuki Nakano,[7] Yasunori Nishimura,[12] Yuji Ogawa,[12] Masayuki Sasaki,[9] Juri Sugimoto,[2] Toshihiro Takagi,[2] Ryuichi Usui,[4] Takayuki Yamamoto,[2] Makoto Yamauchi,[12] AND Koshiro Yoshidome[12]

*Accepted on August 17, 2012*

## ABSTRACT

We present the catalog of high Galactic-latitude ($|b| > 10°$) X-ray sources detected in the first 37-month data of Monitor of All-sky X-ray Image (MAXI) / Gas Slit Camera (GSC). To achieve the best sensitivity, we develop a background model of the GSC that well reproduces the data based on the detailed on-board calibration. Source detection is performed through image fit with the Poisson likelihood algorithm. The catalog contains 500 objects detected in the 4–10 keV band with significance of $s_{D,4-10keV} \geq 7$. The limiting sensitivity is $\approx 7.5 \times 10^{-12}$ ergs cm$^{-2}$ s$^{-1}$ ($\approx 0.6$ mCrab) in the 4–10 keV band for 50% of the survey area, which is the highest ever achieved as an all-sky survey mission covering this energy band. We summarize the statistical properties of the catalog and results from cross matching with the Swift/BAT 70-month catalog, the meta-catalog of X-ray detected clusters of galaxies, and the MAXI/GSC 7-month catalog. Our catalog lists the source name (2MAXI), position and its error, detection significances and fluxes in the 4–10 keV and 3–4 keV bands, their hardness ratio, and basic information of the likely counterpart available for 296 sources.

*Subject headings:* catalogs — surveys — galaxies: active — X-rays: galaxies

## 1. INTRODUCTION

Monitor of All-sky X-ray Image (MAXI: Matsuoka et al. 2009) is the first astronomical mission performed on the International Space Station (ISS). The main instrument of MAXI is the Gas Slit Camera (GSC: Mihara et al. 2011; Sugizaki et al. 2011), which covers the 2–30 keV band. Since its first light on 2009 August 15, MAXI/GSC has been monitoring nearly the entire sky every 92 minute with two instantaneous field-of-views of $1.5° \times 160°$ that rotate according to the orbital motion of the ISS. The survey with MAXI/GSC is expected to achieve the best sensitivity so far as an all-sky monitor in the 2–10 keV band, which is complementary to other all-sky surveys conducted below 2 keV (ROSAT: Voges et al. 1999, 2000) and above 10 keV (INTEGRAL: Bird et al. 2007; Beckmann et al. 2006, 2009; Krivonos et al. 2007; Bird et al. 2010, Swift: Tueller et al. 2008, 2010; Cusumano et al. 2010; Baumgartner et al. 2010, 2012). This bandpass is the most suitable to survey X-ray populations with intrinsically soft X-ray spectra with little bias for obscuration.

Utilizing the first 7-month data of MAXI/GSC in the 4–10 keV band, Hiroi et al. (2011) produced the first X-ray source catalog in the high Galactic latitude sky ($|b| > 10°$). It consists of 143 objects above the $7\sigma$ significance level with a limiting sensitivity of $\sim 1.5 \times 10^{-11}$ ergs cm$^{-2}$ s$^{-1}$ (1.2 mCrab) in the 4–10 keV band. Though limited in sample size, the first MAXI/GSC catalog has an advantage that the identification completeness is very high ($> 99\%$). Utilizing the AGN sample of this catalog, Ueda et al. (2011) calculate a new AGN X-ray luminosity function in the local universe and solve the discrepancy of the AGN space density previously reported from different missions.

In this paper, we present the second MAXI/GSC source catalog at $|b| > 10°$, constructed from the data in the 4–10 keV band observed for 37 months between 2009 September and 2012 October. The significantly increased photon statistics enables us to achieve much better sensitivities than those of the 7-month catalog. We also analyze the data in the 3–4 keV band to derive the hardness ratio of the detected sources. Section 2 describes the data reduction and filtering. In section 3, we report the analysis procedures to detect source candidates and determine their fluxes and positions. Section 4 presents

Electronic address: hiroi@kusastro.kyoto-u.ac.jp

[1] Department of Astronomy, Kyoto University, Oiwake-cho, Sakyo-ku, Kyoto 606-8502
[2] MAXI team, Institute of Physical and Chemical Research (RIKEN), 2-1 Hirosawa, Wako, Saitama 351-0198
[3] ISS Science Project Office, Institute of Space and Astronautical Science (ISAS), Japan Aerospace Exploration Agency (JAXA), 2-1-1 Sengen, Tsukuba, Ibaraki 305-8505
[4] Department of Physics, Tokyo Institute of Technology, 2-12-1 Ookayama, Meguro-ku, Tokyo 152-8551
[5] School of Dentistry at Matsudo, Nihon University, 2-870-1 Sakaecho-nishi, Matsudo, Chiba 101-8308
[6] Department of Physics, Nihon University, 1-8-14 Kanda-Surugadai, Chiyoda-ku, Tokyo 101-8308
[7] Department of Physics and Mathematics, Aoyama Gakuin University, 5-10-1 Fuchinobe, Chuo-ku, Sagamihara, Kanagawa 252-5258
[8] Department of Physics, Chuo University, 1-13-27 Kasuga, Bunkyo-ku, Tokyo 112-8551
[9] Department of Earth and Space Science, Osaka University, 1-1 Machikaneyama, Toyonaka, Osaka 560-0043
[10] Astro-H Project team, Institute of Space and Astronautical Science (ISAS), Japan Aerospace Exploration Agency (JAXA), 3-1-1 Yoshino-dai, Chuo-ku, Sagamihara, Kanagawa 252-5210
[11] National Astronomical Observatory of Japan, 2-21-1, Osawa, Mitaka City, Tokyo 181-8588
[12] Department of Applied Physics, University of Miyazaki, 1-1 Gakuen Kibanadai-nishi, Miyazaki, Miyazaki 889-2192
[13] School of Physical Science, Space and Astronautical Science, The graduate University for Advanced Studies (Sokendai), Yoshinodai 3-1-1, Chuo-ku, Sagamihara, Kanagawa 252-5210
[14] Institute of Space and Astronautical Science (ISAS), Japan Aerospace Exploration Agency (JAXA), 3-1-1 Yoshino-dai, Chuo-ku, Sagamihara, Kanagawa 252-5210
[15] Mitsubishi Space Software Co., Ltd., WTC Bldg.2-4-1 Hamamatsu-cho, Minato-ku, Tokyo 105-6132

the source catalog, its basic properties, the results from cross-matching with other X-ray source catalogs, and the log $N$-log $S$ relation. We summarize our conclusion in section 5.

## 2. DATA REDUCTION

The second MAXI/GSC catalog is based on the 37-month data obtained from 2009 September 23 to 2012 October 15. The data reduction procedure is essentially the same as used to produce the MAXI/GSC 7-month catalog in Hiroi et al. (2011). We start from processed event files provided by the MAXI team, which have columns of the photon arrival time (TIME), energy (Pulse-height Invariant = PI), and sky position (R.A. and Decl.). In the processing, two pulse-height values read out from both ends of a carbon-anode wire were converted into the PI and an one-dimensional detector position along the anode wire on the basis of the ground and in-flight calibration (Mihara et al. 2011; Sugizaki et al. 2011). The sky position of the incident X-ray photon is then calculated by referring to the ISS attitude at the photon arrival time. For the data reduction, we use HEAsoft (version 6.12) and MXSOFT, an original MAXI software package developed by the MAXI team.

To detect sources, the data in the 4–10 keV band are utilized for consistency with our previous work (Hiroi et al. 2011), where high signal-to-noise ratio is achievable due to the high quantum efficiency and low background rate of the detector (Mihara et al. 2011). We basically use all available data of all twelve GSC counters operated with high voltage levels of both 1650 V and 1550 V, except for those taken with GSC_3 after 2010 June 22, in which the background-veto counters were not available. Unlike Hiroi et al. (2011), the photon events of anode #1 and #2 are also available in this study, thanks to the improvement of the instrument response calibration. We discard the data acquired in unusual conditions where the background reproducibility of the GSC is found to be poor: (1) the initial operation phase from 2009 September 1 to 22, (2) the phase from 2012 October 16 to 31 when the Dragon spacecraft was attached to the ISS, (3) those during the docking or undocking operation of the Soyuz spacecraft, (4) those soon after the reboot of the electrical system of MAXI, (5) those during large solar flares[16], and (6) those when the telemetry data were missed due to the network trouble between the ISS and the ground base station.

The following data selection criteria are applied in event selection. To exclude data of the GSC with high particle background, we select the events when the latitude of the ISS is between −40° and 40° and those detected in an inner region of each camera with the incident angle $|\phi| \leq 38°$ (for definition of $\phi$, see Mihara et al. 2011). In our new analysis for the 37-month catalog, we also properly take into account the effects of the occultation of the GSC field-of-views by the space shuttle and the solar paddles of the ISS in a time dependent manner; the occulted regions in the detector coordinate defined by $\phi$ at a given time, if any, are excluded with a margin of 5° on the basis of monitoring information of the relative geometry between MAXI and the ISS.

The left panel of figure 1 displays the effective exposure map for the 37-month MAXI/GSC data in the Galactic coordinates projected with the Hammer-Aitoff algorithm. The effective-exposure distribution is shown in the right panel.

The map is produced by the latest version of the MAXI simulator **maxisim** (Eguchi et al. 2009) for uniformly extended emission, where the $\phi$ dependence of the slit area, the collimator field-of-view, and the detector efficiency as a function of an incident X-ray energy (see Mihara et al. 2011; Sugizaki et al. 2011) are correctly taken into account. Referring to the orbit and attitude of the ISS, **maxisim** is able to generate simulated photon events from any input X-ray sources by reflecting the latest responses of the GSC. The stripe patterns found in the map are caused by an incident-angle dependence of the effective area of the GSC aimed at each pointing direction, the different duty cycle among the twelve GSC counters, and the precession of scanning motion of the field-of-views coupled with that of the ISS orbit. The two darkest, annular-like structures at the top-left and bottom-right of the map correspond to the regions with the longest exposure near the poles around the rotation axis of the ISS orbit.

## 3. ANALYSIS

### 3.1. *Background Reproduction*

For source detection, we perform image analysis of the projected sky image from the whole MAXI/GSC data, as detailed in section 3.2. Basically, we search for excess signals over the profile of the background, which consists of the non-X-ray background (NXB) and the cosmic X-ray background (CXB). Thus, to achieve the best sensitivity, it is critical to model the background image with the least systematic errors that enables us to detect even the faintest sources as statistically significant signals. The NXB of the GSC shows complex behaviors (Sugizaki et al. 2011). The rate of the NXB is highly variable, depending on the environment such as conditions of other spacecrafts attached to the ISS, the orbital position and attitude of the ISS.

We thus construct a background model of MAXI/GSC based on the detailed calibration of the on-board data as described below. Here we do not distinguish the NXB from the CXB, only treating their sum, since it is not easy to separate the two components from the observed data. This is mainly due to the fact that the MAXI/GSC field-of-views always look at the sky and seldom look at the earth. For this purpose, we analyze the on-board data that can be regarded as blank sky, by excluding regions around the 70 brightest sources at high Galactic latitudes listed in the 7-month MAXI/GSC catalog [17] (Hiroi et al. 2011) and that of the Galactic-center region with $|l| < 50°$ and $|b| < 10°$. The same screening described in section 2 is adopted for event extraction. To avoid complexity in exposure correction, we discard any data taken when a part of the field-of-view was occulted by the space shuttle and the solar paddles of the ISS in each camera.

We first investigate the long-term variability of the background rate. The top panel of figure 2 displays a one-day averaged background rate as a function of time, derived from the data in the 4–10 keV band taken with the GSC_4 counter. It is clearly seen that the MAXI/GSC background rate has two separate levels of 3–4 cnts/s and 2–2.5 cnts/s in this camera. We find that this long-term variability of the background is strongly correlated with whether the Soyuz spacecraft was docked or undocked to the ISS. Since Soyuz has an altimeter containing a radioactive source, the MAXI/GSC background rate is largely increased when it is attached to the nearest ISS port to MAXI, and vice versa. Furthermore, it is found that an increment of the background rate depends on the individual Soyuz spacecraft docked at that time. The difference of high-voltage level between 1650 V and 1550 V also affects the background rate. Thus, we divide the 37-month period into 9 different phases according to the existence of Soyuz and the high-voltage level of the GSC, and create the background model separately in each phase.

---

[16] From the light curves with 16-sec bin size in the 4–10 keV band, we filter out time regions when the count rates exceeded 25, 35, 40, and 30 cnts/s for GSC_3, GSC_8, GSC_B, and the other cameras, respectively, with margins of 32 sec each before and after these periods.

[17] We confirm that when our new 37-month catalog is used to exclude even fainter sources in this process, the resultant background profile in the detector coordinate is little ($< 1\%$) changed.

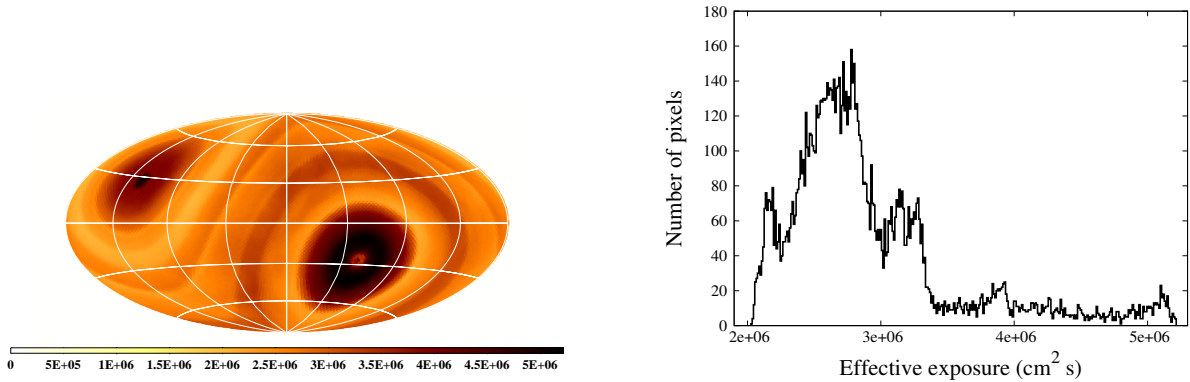

FIG. 1.— (Left) Effective exposure map for the 37-month MAXI/GSC data in the Galactic coordinates projected with the Hammer-Aitoff algorithm. The units are cm$^2$ s. The entire sky is divided into 12288 pixels with a size of $1.83° \times 1.83°$. (Right) Histogram of the effective exposure presented in the left panel.

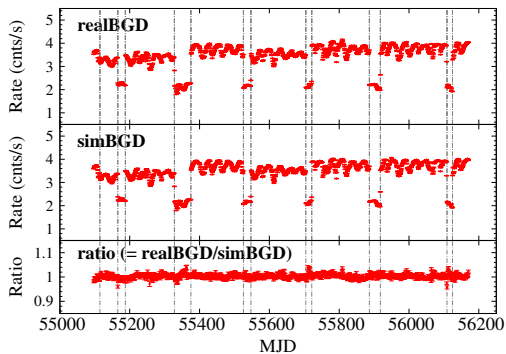

FIG. 2.— (Top) History of the background rate in the 4–10 keV band observed with GSC_4 in a one-day bin (real data). The attached error is $1\sigma$. The dot-dashed lines indicate the boundaries of the phases. The dock/undock status of the Soyuz spacecraft causes the drastic change of the daily background rate. (Middle) Same as the top panel but produced from simulations. (Bottom) The ratio of the real background rate to the simulated one. The simulated background data are well consistent with the real ones with an rms scatter of $\sim 1\%$.

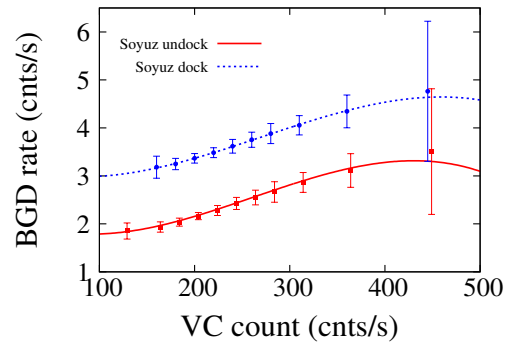

FIG. 3.— Correlation between the 4–10 keV background rate and the VC count rate for GSC_4. The red squares (for visuality, slightly shifted) and solid curve represent the data points and the best-fit model of a third-order polynomial, respectively, when Soyuz was undocked, while the blue circles and dotted curve show them when it was docked from 2009 December 22 to 2010 May 11 (MJD = 55187–55327).

Next, the short-term variability of the NXB on a time scale less than one day is examined. In the previous study (Hiroi et al. 2011), we considered the dependence of the background rate on the cut-off rigidity (COR) parameter, which is determined solely by the orbital position of the ISS. The correlation of the NXB with COR is known to be not so good, however. Alternatively, we have found that the instantaneous background rate is well estimated from the so-called "VC" count, representing the rate of events simultaneously detected by the carbon-anode cells and the tungsten-anode cells for particle veto, available as a monitor count in the house keeping data of the GSC (Mihara et al. 2011; Sugizaki et al. 2011). The reason why the VC works as a good monitor of the NXB is that high energy particles hitting the GSC counter should be detected by both the inner (carbon) and outer (veto) cells, unlike X-ray events. Figure 3 plots examples of the correlation between the VC and background rates in the 4–10 keV band for GSC_4 in two different phases. The blue circles and red squares correspond to the data obtained when Soyuz was docked to the ISS port and undocked, respectively. The error bars represent $1\sigma$ standard deviation in the background rate including the statistical fluctuation. A tight correlation is confirmed in each phase. The blue-dotted and red-solid lines represent the best-fit models of a third-order polynomial to these correlations.

We develop a background-event generator of the GSC, which makes use of the framework of the MAXI simulator **maxisim**. By using the correlation between the VC and background rates determined for each camera and phase, it generates simulated background events in time series with the instantaneous rate calculated from the light curve of the VC counts. Their energies and positions on the detector are randomly assigned according to their distributions derived from the real background data, which also depend on the VC count. We produce ten times more events than the real data to reduce the statistical error in the model. The simulated background data are then processed in the same way as for the observed data. The middle panel of figure 2 plots the same as that shown in the top panel of figure 2 but based on the simulated background data. We confirm that the background light curve is well reproduced by our model with an rms scatter of $\sim 1\%$, as shown in the bottom panel of figure 2.

### 3.2. *Image Analysis*

In this section, we describe the details of our image analysis method to produce the X-ray catalog from the MAXI/GSC

data. The procedure consists of two major stages: a detection of source candidates from smoothed images, and a measurement of fluxes and positions for the source candidates by maximum-likelihood image fitting. For the image analysis, we use display45 version 2.04, which is a command-driven, interactive histogram-browsing program based on CERN libraries.

### (1) MAKING PARTED IMAGES

Both for the real and simulated background data, we create images from the event lists after applying the selection criteria described in section 2. Like a standard image analysis, we produce a tangentially-projected image within a small region in the sky coordinates $(X, Y)$ defined by three parameters: the reference point in right ascension (R.A.) and declination (Decl.) corresponding to the center of the image and the position angle of the $+Y$-axis to the north, which is set to be zero in our analysis. The images are simply stored in units of counts without exposure correction.

The entire sky is divided into 768 square regions with a size of $7.3° \times 7.3°$, whose centers are determined by using the HEALPix software package (Górski et al. 2005) so that each region has the same solid angle. Using the central position of each region as the sky-reference point, we then make 768 tangentially-projected images with a size of $14° \times 14°$ by referring to the columns of sky position (R.A., Decl.) in the event lists. In order not to miss any sources located near the boundaries, the images have overlapping area of $> 3°$ on both sides with neighboring ones. The bin size is set to $0.1°$, which is sufficiently finer than the typical size of the point spread function (PSF) of MAXI/GSC (FWHM$\sim 1.5°$).

### (2) SEARCHING FOR SOURCE CANDIDATES

To pick up source candidates from each image, we create a significance map in the following manner. Firstly, we make net-source maps by subtracting the simulated background from the real data. Then, both the real data and net-source map are smoothed with a circle of $r = 1°$ with a constant weight of unity (i.e., simple integration). We finally calculate the significance of excess signals as "net-source/$\sqrt{\text{real data}}$" at each point, yielding the significance map.

In this step, a correct estimate of the background level is essential. Although the reproductivity of the background model described in section 3.1 is fairly good for blank sky, we find that it is sometimes better to re-normalize the absolute level of the background map from the observed data themselves, in particular when additional diffuse background other than the CXB such as Galactic diffuse emission exists in the image. Thus, the background level used to make a final source-candidate list is determined by iteration in the following way. At first, assuming a conservative background level higher than the nominal value by a factor of 1.1, we create a tentative significance map. We then search for the peak showing the highest significance in the map if it is larger than $15\sigma$. Next, after masking out the circular region around the peak of the bright source with a radius of $r = 3°$, the whole size of the PSF, we compare the averaged counts in the real and the simulated background maps, whose ratio determines an improved background level. By repeating these procedures of finding and masking out brightest sources until the significance of a newly detected peak becomes $< 15\sigma$, we finally derive the most reasonable background level in each image. The correction from the nominal level is typically less than 3%, and thus possible systematic errors in the positional dependence of the background do not affect the flux measurements of sources.

Once the background level is determined, we make the most reliable significance map, from which peaks above $5\sigma$ are picked up as source candidates. Those located near the boundaries of the image ($|X| > 5.5°$ or $|Y| > 5.5°$) are discarded because they are located closer to the center of a neighboring image that can detect them with better conditions. The issue of source confusion that two nearby sources may be detected as a single source candidate will be overcome in a later procedure (the 6th step).

### (3) MERGING SOURCE-CANDIDATE LIST

Merging the source-candidate lists obtained from the whole sky yields 1308 sources in total above $5\sigma$ at $|b| > 10°$. This list includes duplicate sources detected in multiple images, however, which are located at the overlapping regions near the boundaries. We treat any pair of sources whose positions are within $1°$ as a single object; we adopt a source detected at a closer position to the image center and discard the other. It is confirmed that the detection significances of the paired candidates are consistent with each other within expected levels. Thus, we finally obtain a unique source-candidate list consisting of 615 objects above $5\sigma$ at $|b| > 10°$.

### (4) CONSTRUCTING MODELS OF POINT SPREAD FUNCTION

For image analysis, we need to construct a model of the PSF for each source candidate. Since the MAXI/GSC data used in this study compose of multiple observations with different conditions, it is difficult to model the PSF shape analytically. Hence, we utilize the MAXI simulator **maxisim** (Eguchi et al. 2009) to produce the PSF model by inputting a point-like source at the detected position. The simulation is performed with exactly the same observational conditions (such as the total exposure) as for the real data, and hence the resultant PSF contains the expected number of counts from that source with a given flux by fully taking into account the instrumental responses. The Crab Nebula-like spectrum[18], a power law of a photon index of 2.1 absorbed with a hydrogen column density of $N_{\rm H} = 2.6 \times 10^{21}$ cm$^{-2}$, and no flux variability are assumed for simplicity. We confirm that the choice of the spectrum for the PSF model does not affect the determination of the flux by the image fitting because the energy dependence of the PSF is not very large in the energy band of interest. To suppress the statistical fluctuation in the PSF model, we generate a sufficiently large number of photons for each source candidate by assuming a flux of 50 mCrab. The simulated events of the PSF data are processed in the same way as for the real data, and are converted into images in the sky coordinates.

### (5) DETERMINING FLUXES AND POSITIONS WITH IMAGE FIT

To determine the fluxes and positions of the source candidates, we perform image fitting to the real data with a model consisting of the background and PSF models, both produced by the simulations as described above. Here the fits to all of the source candidates and the background are performed simultaneously in each image. The best-fit parameters are determined with the Poisson likelihood algorithm where the

---

[18] The spectral parameters are adopted from the INTEGRAL General Reference Catalog (ver. 35). http://www.isdc.unige.ch/integral/science/catalogue

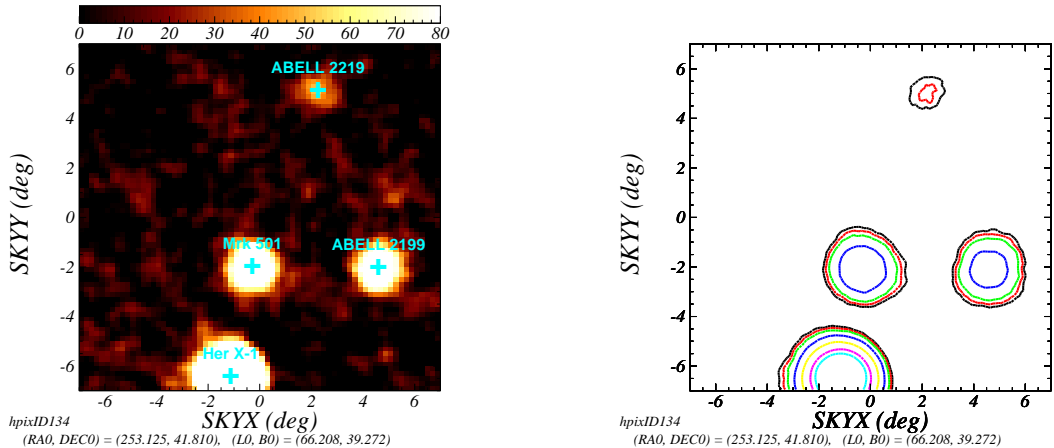

FIG. 4.— (Left) Example of a smoothed image of the real data. The image size is $14° \times 14°$, whose center is located at (R.A., Decl.) = (253.125°, 41.810°). The best-fit background model obtained from image fitting is subtracted. The four cyan crosses represent the locations of detected sources with $s_{\rm D,4-10keV} \geq 7$. (Right) Significance map of the same region. The contours display the significance levels of 5 (black), 7 (red), 10 (green), 20 (blue), 40 (yellow), 70(magenta), and 90(cyan), from outer to inner.

so-called $C$ statistics (Cash 1979) is minimized, defined as

$$C \equiv 2 \sum_{i,j} \{M(i,j) - D(i,j) \ln M(i,j)\}, \quad (1)$$

where $D(i,j)$ and $M(i,j)$ represent the data and model at the image pixel $(i,j)$, respectively. The parameter that gives a $C$ value larger by unity than that of the best-fit with the other free parameters allowed to float corresponds to its $1\sigma$ error. The MINUIT package is utilized in the minimization. In the fitting, we change the bin width of an image from 0.1° to 0.2° for the fit to converge faster, and only treat the inner $11° \times 11°$ region of the image for the calculation to ignore the influence by sources located just outside the whole image with a size of $14° \times 14°$. For the PSF models of source candidates located in the inner $11° \times 11°$ region of the image, their normalizations and positions are set to be free, as well as the overall background level. The PSF normalization of each source gives its flux corrected for exposure and the instrumental response in units of "Crab" because we basically measure its total counts relative to that would be observed from the Crab Nebula on which the PSF simulation is based. Since it is difficult to accurately determine the position of a source outside the $11° \times 11°$ region whose PSF is distributed beyond the image boundary, we fix it at the location originally detected in the significance map. It is found that in a few images containing very bright sources such as Sco X-1 and Cyg X-2, the background level cannot be properly determined due to the incomplete reproduction of the PSF shape. In such cases, we fix the background level to the value obtained in the second step.

After the fitting, we calculate the detection significance in the 4–10 keV ($s_{\rm D,4-10keV}$) for each source candidate, defined as

$$s_{\rm D,4-10keV} \equiv (\text{best-fit 4-10 keV flux})/(\text{its } 1\sigma \text{ statistical error}). \quad (2)$$

Here we use the MINOS negative error in the MINUIT package as the flux error, although the positive and negative errors are consistent with each other in almost all cases because the number of source counts is always sufficiently high that symmetric errors are a reasonable approximation. We adopt $s_{\rm D,4-10keV} \geq 7$ for the criterion of source detection, considering the number of fake sources caused by the fluctuation of background level (see section 4.2).

*(6) ITERATION OF SOURCE SEARCH AND IMAGE FIT*

Due to source confusion, multiple close sources can be detected as a single object. To salvage such sources located in a source crowded region as much as possible, we again perform the source search in the second step but by adopting the best-fit total model consisting of PSFs and the background obtained in the previous step as a new background. Here we adopt a more conservative detection criterion of $5.5\sigma$ than in the first round to reduce the probability of fake detections around the existing sources. This procedure gives us an additional source-candidate list containing 135 objects above $5.5\sigma$ at $|b| > 10°$. The first and second source-candidate lists are merged into one list. We then repeat the image fitting using the revised source-candidate list, and obtain the final list of sources detected with $s_{\rm D,4-10keV} \geq 7$ in the 4–10 keV band.

*(7) MEASUREMENT OF 3–4 KEV FLUX*

To obtain spectral information, we also determine the 3–4 keV fluxes of all sources in the final list created above. The images of real data, background model, and PSF models in the 3–4 keV band are produced in the same way as for those in the 4–10 keV band. We then perform image fitting to the 3–4 keV band image using the source-candidate list obtained in the 6th step. The source positions are fixed at those determined by the image fit in the 4–10 keV band. Thus, we finally derive the 3–4 keV fluxes and their errors for all the sources detected in the 4–10 keV band. Using the 4–10 keV and 3–4 keV fluxes, which are independently determined from the images of the two energy bands, we calculate the hardness ratio (HR) defined as

$$\text{HR} = \frac{H-S}{H+S}, \quad (3)$$

where $H$ and $S$ represent the observed fluxes in the 4–10 keV and 3–4 keV bands, respectively. The error of the HR is calculated by simple propagation of those in $H$ and $S$, both can be regarded as symmetric errors.

## 4. CATALOG

### 4.1. *Basic Properties*

In this section, we present the second MAXI/GSC X-ray source catalog in the high Galactic latitude sky obtained from the 37-month all-sky survey data in the 4–10 keV band.

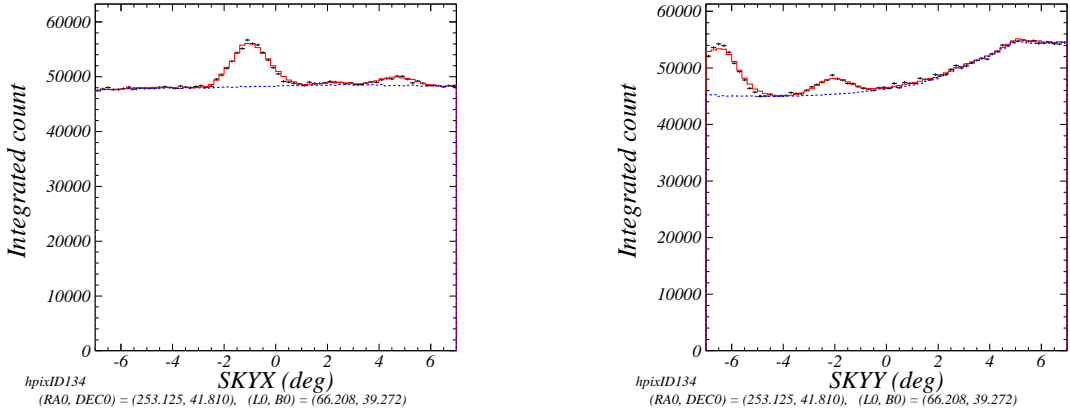

FIG. 5.— Projected images onto the *X*-axis (left) and *Y*-axis (right) of the same region as shown in figure 4. The black points represent the real data with $1\sigma$ Poisson errors. The two color lines display the best-fit models (red-solid: total, blue-dashed: only background).

It consists of 500 sources detected with a significance of $s_{D,4-10\text{keV}} \geq 7$ at $|b| > 10°$. Table 1 gives the catalog with columns of (1) source identification number, (2) MAXI source name based on the best-fit position in the final image fit (with a prefix of "2MAXI"), (3)-(4) best-fit position in the equatorial coordinates in units of degree, (5) $\sigma_{\text{stat}}$: $1\sigma$ statistical position error in units of degree (the squared sum of those in the X- and Y-directions). (6) $s_{D,4-10\text{keV}}$: 4–10 keV detection significance, (7) $f_{4-10\text{keV}}$: observed 4–10 keV flux and its $1\sigma$ error in units of $10^{-12}$ ergs cm$^{-2}$ s$^{-1}$, (8) $s_{D,3-4\text{keV}}$: 3–4 keV detection significance, (9) $f_{3-4\text{keV}}$: observed 3–4 keV flux and its $1\sigma$ error in units of $10^{-12}$ ergs cm$^{-2}$ s$^{-1}$, (10) HR: hardness ratio and its $1\sigma$ error, (11) counterpart name, (12)-(13) position of the counterpart (14) redshift, (15) type, (16) cross-matching flag, and (17) note.

Taking into account the MAXI/GSC response, we calculate the dependences of the hardness ratio HR on spectral parameters and those of the conversion factors from an area-corrected count rate (defined as that relative to the Crab one) into an energy flux in the same energy band, $C_{4-10}$ and $C_{3-4}$. In the top-left panel of figure 6, the hardness ratio is plotted as a function of photon index with an unabsorbed power-law spectrum, whereas in the top-right panel it is given as a function of column density of an absorbed power-law spectrum with a photon index of 1.8 at redshift of 0. In the middle and bottom panels, we also display the dependences of the conversion factors on the same spectral parameters in the 4–10 keV and 3–4 keV bands, respectively. The expected values of HR and conversion factors are HR=0.0745, $C_{4-10} = 1.24$ [$10^{-11}$ ergs cm$^{-2}$ s$^{-1}$]/[mCrab], $C_{3-4}$ =0.401 [$10^{-11}$ ergs cm$^{-2}$ s$^{-1}$]/[mCrab] for a photon index of 1.8 with no absorption, and HR=0.000, $C_{4-10}$ =1.21 [$10^{-11}$ ergs cm$^{-2}$ s$^{-1}$]/[mCrab], $C_{3-4}$ =0.398 [$10^{-11}$ ergs cm$^{-2}$ s$^{-1}$]/[mCrab] for the Crab Nebula-like spectrum (see section 3.2 for its spectral parameters). The energy fluxes listed in the catalog are all calculated by assuming the Crab Nebula-like spectrum.

The correlation between the flux and the detection significance in the 4–10 keV band are plotted in figure 7. The detection significance, $s_D$, can be approximated as $St/\sqrt{St+Bt}$, where $S$ and $B$ are the count rates of the source and background, respectively, and $t$ is exposure. Thus, in the high flux range ($s_D \sim \sqrt{St}$), the detection significance is expected to be proportional to (flux)$^{1/2}$. In contrast, it is proportional to (flux)$^1$ in the low flux range ($s_D \sim S\sqrt{t/B}$). These trends are confirmed in the figure. The limiting sensitivity with $s_{D,4-10\text{keV}} \geq 7$ (that of the faintest source detected) corresponds to $\sim 5 \times 10^{-12}$ ergs cm$^{-2}$ s$^{-1}$ ($\sim 0.4$ mCrab) in the 4–10 keV band.

### 4.2. *Number of Spurious Sources*

Since we conservatively adopt the detection threshold of $s_D \geq 7$, fake detections by statistical fluctuation in photon counts are expected to be negligible. However, the catalog may contain a small fraction of spurious sources caused by remaining systematic errors in the background model. To simply estimate the number of such sources, we search for "negative" signals from the smoothed net-source maps produced in the second step of section 3.2. This gives a "negative" source-candidate list consisting of 50 objects below $-7\sigma$ from the entire sky. Performing image fitting by adding negative PSFs to the model, we find that the number of final "negative" detections with $s_D \leq -7$ becomes only 18% of that of spurious source candidates. The reduction is expected because the shape of residual signals due to improper modeling of the background profile is very different from that of the PSF. Thus, we estimate that our catalog would include $\sim 9$ spurious detections, which are less than 2% of the total sources. Note that if we instead adopt $s_D \geq 6$ as the detection criterion for the catalog, we find that the fraction of spurious ones becomes $\sim 5\%$ of the total, which is too high to be accepted.

### 4.3. *Cross Matching with Other X-ray Catalogs*

Identification of X-ray sources with counterparts in other wavelengths is very important for further scientific studies using an X-ray catalog. The most critical information for this task is the position accuracy. Figure 8 plots the correlation between the flux and position error of the MAXI sources. As noticed, $\sigma_{\text{stat}}$ is roughly proportional to $(s_{D,4-10\text{keV}})^{-1}$, as expected from a simple statistical argument on the accuracy of the centroid determination of the PSF with limited photon counts. A typical $1\sigma$ error becomes $\sigma_{\text{stat}} \sim 0.2°$ at $\sigma_{D,4-10\text{keV}} = 7$.

Considering the large positional uncertainties of MAXI, it is difficult to directly match them with those in optical/infrared catalogs with large number densities. However, since the MAXI sources represent brightest X-ray populations, a significant fraction of them are expected to be included by other X-ray catalogs at similar flux levels covering a wide sky area. Hence, as the first step of identification work, we cross-match our sources with two major X-ray catalogs to determine the most likely counterparts: the Swift/BAT hard X-ray (14–195 keV) all-sky survey 70-month catalog (hereafter, BAT70; Baumgartner et al. 2012) and the Meta-Catalog of X-Ray Detected Clusters of Galaxies (hereafter, MCXC; Piffaretti et al. 2011). The latter is based on the ROSAT all-sky survey and ROSAT serendipitous surveys performed in the energy bands below ∼2 keV. While BAT70 provides an X-ray sample including heavily obscured objects, MCXC complements the identification of galaxy clusters, which are hard to be detected above 10 keV due to their soft spectra.

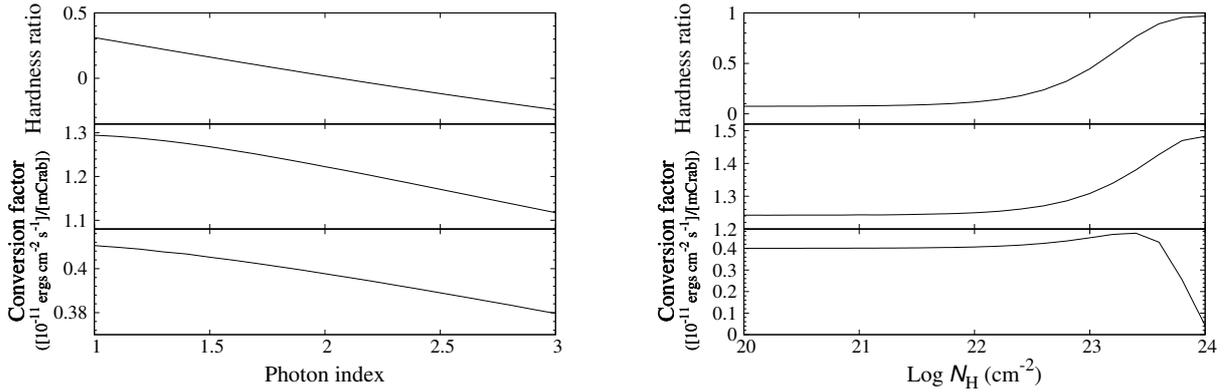

FIG. 6.— (Top) Hardness ratio between the 4–10 keV and 3–4 keV bands (HR) given as a function of a spectral parameter. (Middle) The conversion factor from a count rate (defined as that relative to the Crab one) to an energy flux in the 4–10 keV band for different spectral parameters in units of [$10^{-11}$ ergs cm$^{-2}$ s$^{-1}$]/[mCrab]. (Bottom) The same as above but in the 3–4 keV band. (Left) The parameter is the photon index of a power law with no absorption. (Right) The parameter is the absorption column density for a power law with a photon index of 1.8.

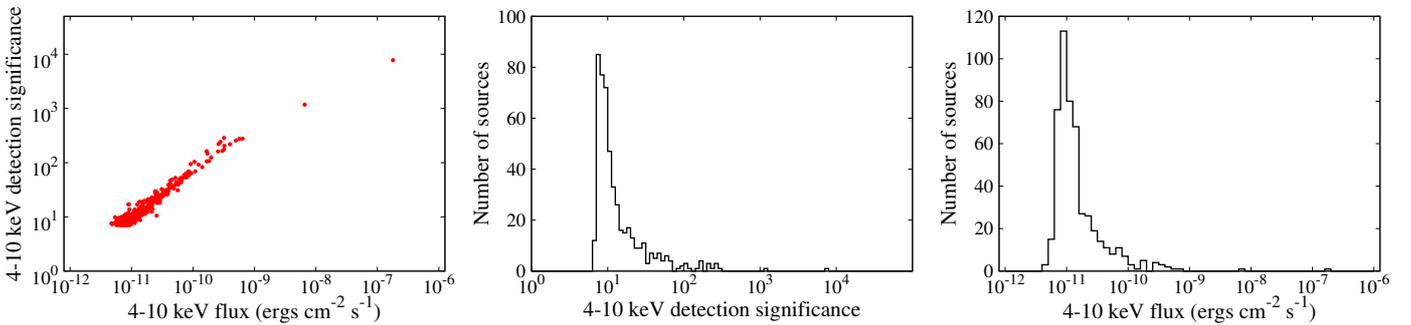

FIG. 7.— Correlation between the flux and the detection significance in the 4–10 keV band (left), histogram of the 4–10 keV flux (middle), and that of 4–10 keV detection significance (right) for all the sources in the catalog.

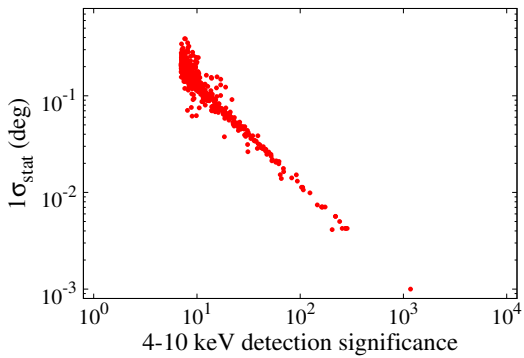

FIG. 8.— Correlation between the 4–10 keV detection significance ($s_{\rm D,4-10keV}$) and the $1\sigma$ positional error ($\sigma_{\rm stat}$).

The cross-matching of the MAXI sources with these catalogs is carried out in the following way. We firstly search for counterparts from the BAT70 catalog. Then, for those without BAT counterparts, we use the MCXC. Finally, we also perform cross-matching with the first MAXI/GSC source catalog (hereafter, GSC7; Hiroi et al. 2011). Basically, we refer to the positions of the optical/infrared counterparts in these catalogs instead of the X-ray positions, except for a few sources in the first MAXI/GSC catalog that do not have unique optical counterparts. As the position errors of the MAXI sources, we consider both statistical error ($\sigma_{\rm stat}$) and systematic error ($\sigma_{\rm sys}$) related to the instrument calibration and attitude determination of MAXI. Since the two errors are independent of each other, we define the total $1\sigma$ position error ($\sigma_{\rm posi}$) as the root sum squares:

$$\sigma_{\rm posi} \equiv \sqrt{\sigma_{\rm stat}^2 + \sigma_{\rm sys}^2}. \quad (4)$$

In the matching process, we set the $1\sigma$ systematic error as $\sigma_{\rm sys} = 0.05°$ on the basis of the previous work (Hiroi et al. 2011), and use an error circle with a radius of $r = 2.5\sigma_{\rm posi}$, which corresponds to 95% confidence level in the 2-dimensional space giving a typical increment in the $C$ value by 6.0. The position errors in the reference catalogs are ignored, since they are much smaller than those of the MAXI sources.

The cross-matching results are listed in columns (11)–(15) of table 1 for each source. The locations of the MAXI sources in the Galactic coordinates are plotted in figure 9. The numbers of matched MAXI sources in a single or two reference

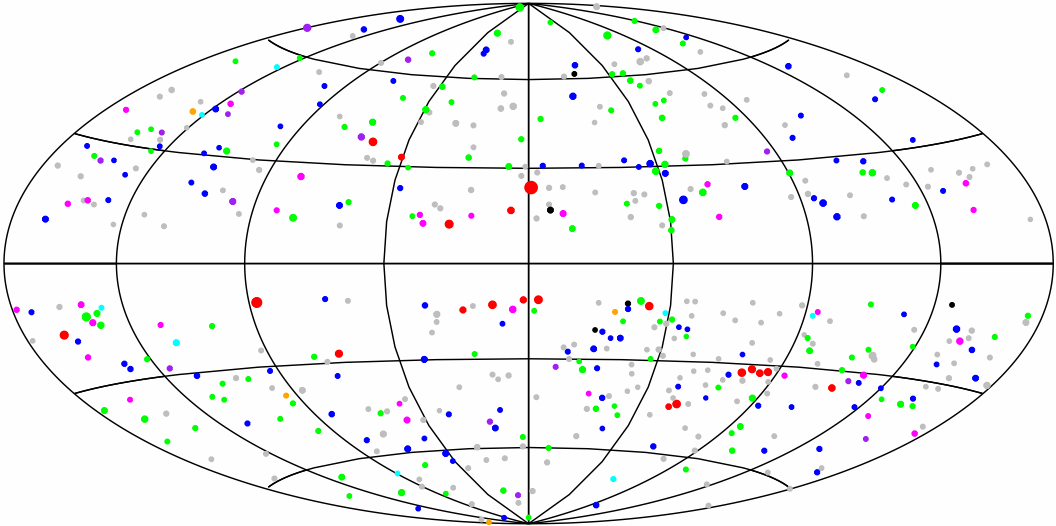

FIG. 9.— Locations of all the MAXI sources in the Galactic coordinates. The center is at $(l,b) = (0°, 0°)$. The radii of the circles are proportional to the logarithm of the fluxes. Different colors correspond to different types: unidentified X-ray sources in BAT70 (black); galaxies (cyan); galaxy clusters (green); Seyfert galaxies, blazars (purple); CVs/Stars (magenta); X-ray binaries (red); confused sources (orange); sources without counterparts in BAT70 or MCXC (gray).

TABLE 2
NUMBER OF MATCHED MAXI SOURCES.

| Catalog name | BAT70 | MCXC | GSC7 | One catalog only[a] |
|---|---|---|---|---|
| BAT70 | 185 (187)[b] | 23 | 84 | 92 |
| MCXC | ..... | 123 (130)[b] | 37 | 78 |
| GSC7 | ..... | ..... | 114 (114)[b] | 8 |

NOTE.— Numbers of MAXI sources matched with both catalogs are listed.
[a] Numbers of MAXI sources matched only with one catalog given in the first column.
[b] Numbers in parentheses represent those of total matched counterparts in each catalog.

TABLE 3
CATEGORIES OF CATALOGED SOURCES.

| Category | Number of sources |
|---|---|
| unidentified | 5 |
| galaxies | 8 |
| galaxy clusters | 114 |
| Seyfert galaxies | 100 |
| blazars | 15 |
| CVs/Stars | 30 |
| X-ray binaries | 20 |
| confused | 4 |
| unmatched | 204 |

catalogs in any combination from the three reference catalogs are summarized in table 2. We also show the total numbers of counterparts in each reference catalog with parentheses; here we count more than one objects in the same reference catalog if they are located within the error radius of the MAXI source. In such cases, we adopt the one with the smallest angular separation from the MAXI position as the most likely counterpart used for the definition of source type. As a result, we find that 296 out of the 500 sources in the present catalog have one or more counterparts. The numbers of each source type are as shown in table 3: 2 unidentified X-ray sources in BAT70, 8 galaxies, 114 galaxy clusters, 100 Seyfert galaxies, 15 blazars, 30 CVs/Stars, 20 X-ray (neutron/black-hole) binaries, 4 confused objects, and 204 sources without counterparts in BAT70 or MCXC.

In the above procedure, some of the MAXI sources may be matched to those in the BAT70 catalog and MCXC by chance. Here we estimate the number of such coincidental match, $N_{\rm CM}$, calculated as

$$N_{\rm CM} = \rho_{\rm BAT70} \times S_{\rm BAT70} + \rho_{\rm MCXC} \times S_{\rm MCXC}, \quad (5)$$

where $\rho_{\rm BAT70(MCXC)}$ and $S_{\rm BAT70(MCXC)}$ represent the source number densities of BAT70 (MCXC) at $|b| > 10°$ and the total areas of the error circles of the MAXI sources used for the cross-matching with those catalogs. Specifically, we use the following values: $\rho_{\rm BAT70} = 0.025$ (deg$^{-2}$), $\rho_{\rm MCXC} = 0.050$ (deg$^{-2}$), $S_{\rm BAT70} = 236.8$ (deg$^2$), and $S_{\rm MCXC} = 173.9$ (deg$^2$). We thus estimate $N_{\rm CM} = 10.3$, which is ∼3% of the total number of the matched sources.

### 4.4. Position Accuracy

We here examine the actual positional error of MAXI as a function of detection significance in the 4–10 keV band on the basis of the cross matching result. We calculate the angular separation between the MAXI position and that of the most likely counterpart for each object. Figure 10 shows the histograms of the angular separation for all the matched sources in different significance bins (left: $s_{\rm D,4-10keV} = 7-12$, middle: $\sigma_{\rm D,4-10keV} = 12-80$, and right: $s_{\rm D,4-10keV} > 80$). Figure 11 plots 90% error radii as a function of 4–10 keV detection significance that are directly obtained from the observed histograms. The black crosses represent those in the bins of $s_{\rm D,4-10keV} =$7-9, 9-12, 12-25, 25-80, and $> 80$, from left to right. We fit these data with the form of

$$\sigma_{\rm posi}^{90\%} = \sqrt{(A/s_{\rm D,4-10keV})^B + C^2}, \quad (6)$$

where $A$, $B$, and $C$ are free parameters. The blue curve in figure 11 represents the best-fit with $A = 2.93\pm0.47$, $B = 1.80\pm0.22$, and $C=0.09\pm0.01$. The slope $B$ is close to 2.0 as expected from the theory. From this result, we estimate the 90% systematic error to be $\approx 0.09°$, which is consistent with that derived in the previous study (Hiroi et al. 2011).

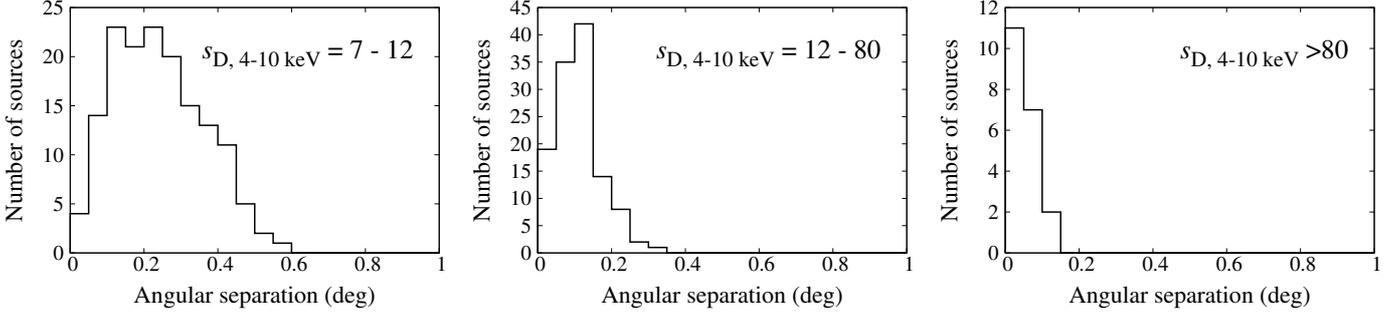

FIG. 10.— Histograms of the angular separation between the positions of the MAXI sources and those of their most-likely counterparts in the reference catalogs. The left, center, and right panel include the objects with the significance level of $s_{D,4-10\rm keV}$ = 7–12, 12–80, and >80, respectively.

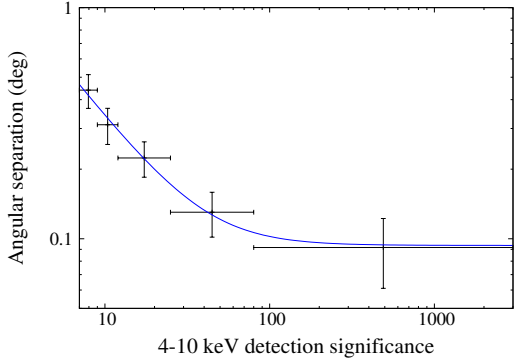

FIG. 11.— 90% error radius of MAXI sources as a function of detection significance in the 4–10 keV band. The black points denote the data and the blue curve represents the best-fit model with the form of $\sqrt{(A/s_{D,4-10\rm keV})^B + C^2}$, where $A = 2.9\pm0.5$, $B = 1.8\pm0.2$, and $C=0.09\pm0.01$.

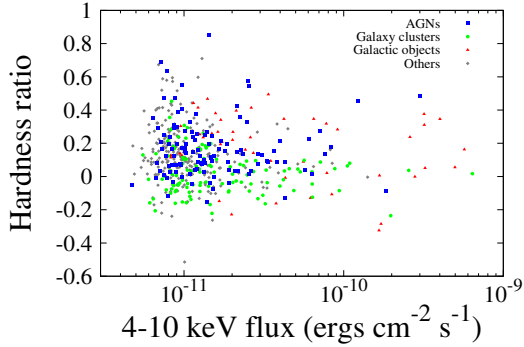

FIG. 12.— Flux in the 4–10 keV band versus HR plot for the cataloged sources with different types: AGNs (blue squares), galaxy clusters (green circles), Galactic/LMC/SMC objects (red triangles), and others (i.e., sources without unique identification; gray diamonds). The typical errors of the HR at $f_{4-10\rm keV}$ of $10^{-10}$ and $10^{-11}$ ergs cm$^{-2}$ s$^{-1}$ are ∼0.01 and ∼0.09, respectively. For visuality, two points with the largest fluxes of $f_{4-10\rm keV} > 10^{-9}$ cm$^{-2}$ s$^{-1}$, corresponding to Sco X-1 and Cyg X-2, are excluded from this figure.

### 4.5. *Hardness Ratio Distribution*

Figure 12 plots the diagram between the 4–10 keV flux and hardness ratio (HR) for the cataloged sources with different types. The blue squares and green circles represent AGNs and galaxy clusters, respectively. The red triangles correspond to Galactic objects and large/small Magellanic cloud (LMC/SMC) objects. Sources without unique counterparts based on the catalog matching described above (i.e., those with no counterpart or multiple ones) are denoted as the gray diamonds. Figure 13 displays the histograms of HR for each type. The *HR* distribution of galaxy clusters is located at lower values than that of AGNs. This is consistent with the expectation that galaxy clusters have softer spectra than AGNs in the 3–10 keV range.

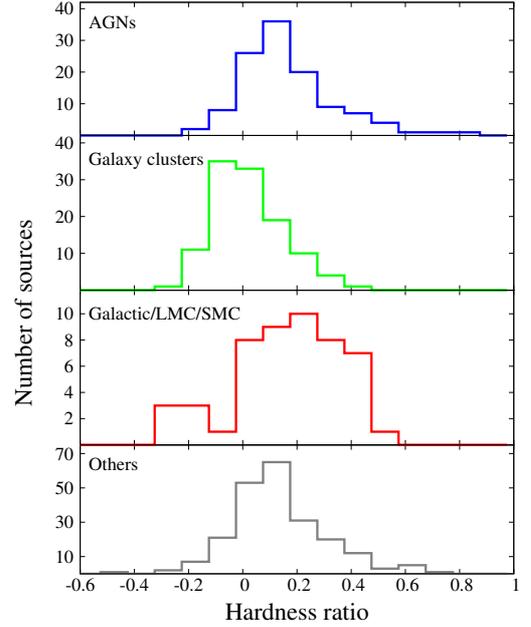

FIG. 13.— Histograms of HR shown in figure 12 for each type: from top to bottom, AGNs, galaxy clusters, Galactic/LMC/SMC, and others.

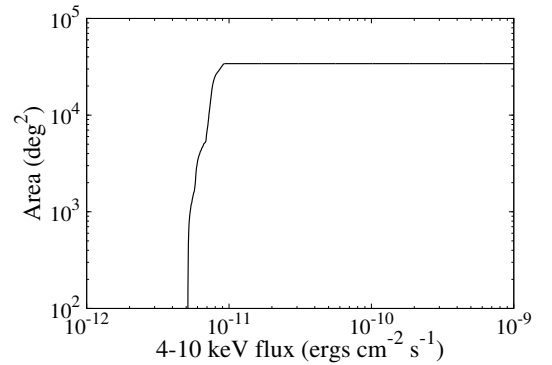

FIG. 14.— Area curve with detection significance of $s_{D,4-10\rm keV} > 7$ plotted against 4–10 keV flux for the MAXI/GSC 37-month survey at $|b| > 10°$.

### 4.6. *Log N-log S Relation*

10                                                                                                              K. Hiroi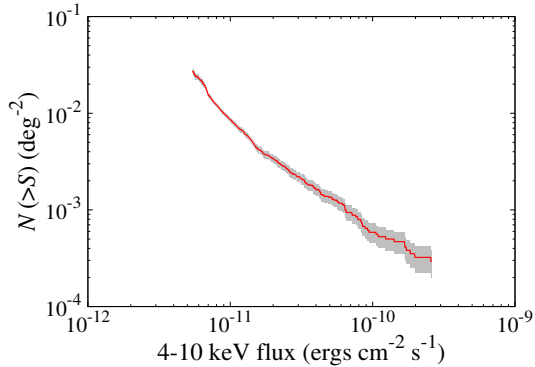

FIG. 15.— Log $N$-log $S$ relation in the 4–10 keV band derived from the MAXI/GSC 37-month survey at $|b| > 10°$. All the 500 sources are included in the calculation. The gray-shaded area represents the $1\sigma$ statistical error region calculated from the source counts.

We derive the source number counts (log $N$-log $S$ relation) in the 4–10 keV band based on the MAXI/GSC 37-month catalog. For this calculation, we need the so-called area curve, which represents a survey area guaranteed for detection of a source given as a function of flux. In the same way as in Hiroi et al. (2011), we estimate the sensitivity at each sky position on the basis of the background photon counts and "effective" exposure (i.e., exposure × detector area). Figure 14 displays the area curve in the 4–10 keV band for the detection significance of $s_{D,4-10\text{keV}} > 7$. As noticed from the figure, the sensitivity at the deepest exposure reaches to $\sim 0.4$ mCrab ($\sim 5 \times 10^{-12}$ ergs cm$^{-2}$ s$^{-1}$ in the 4–10 keV band), which is consistent with the faintest flux level of the cataloged sources.

Dividing the flux distribution of the detected sources by the survey area, we obtain the log $N$-log $S$ relation in the differential form. Figure 15 shows the log $N$-log $S$ relation of all the MAXI sources in the integral form, where the total number density $N(> S)$ for sources with fluxes above $S$ is plotted. The results at fluxes above $1.5 \times 10^{-11}$ ergs cm$^{-2}$ s$^{-1}$ perfectly match with those of Hiroi et al. (2011) derived from the 7-month data. In the faintest range below $7 \times 10^{-12}$ ergs cm$^{-2}$ s$^{-1}$, however, the number density is slightly (by $\sim 10\%$) larger than that extrapolated from the brighter flux range. This is most probably attributable to the effects by source confusion, which is unavoidable at faintest fluxes close to a confusion limit determined by the PSF size. The effects may have to be taken into account in future studies using this catalog, depending on the scientific goals.

## 5. CONCLUSION

We present the second MAXI/GSC source catalog in the high Galactic-latitude sky based on the 37-month data in the 4–10 keV band observed from 2009 September to 2012 October. The accuracy of the background model is significantly improved from our previous work (Hiroi et al. 2011) by considering the dependence of the background rate on the "VC" monitor count sensitive to high energy particles. Through the image fitting to tangentially-projected images with the Poisson likelihood algorithm, we finally detect 500 X-ray sources at $|b| > 10°$ with the detection significance $s_{D,4-10\text{keV}} \geq 7$. The limiting sensitivity is $\sim 0.6$ mCrab, or $7.5 \times 10^{-12}$ ergs cm$^{-2}$ s$^{-1}$ (4–10 keV) for 50% of the survey area, which is the highest ever achieved among all-sky missions covering this energy band. Performing cross-matching with other X-ray source catalogs, we find that 296 sources have one or more than one counterparts. The identification work of a complete sample from this catalog is on-going, which will be reported in future papers.

This work makes use of the HEALPix (Górski et al. 2005) package. This work was partly supported by the Grant-in-Aid for JSPS Fellows for young researchers (KH), Scientific Research 23540265 (YU), and by the grant-in-aid for the Global COE Program "The Next Generation of Physics, Spun from Universality and Emergence" from the Ministry of Education, Culture, Sports, Science and Technology (MEXT) of Japan.

TABLE 1
PROPERTIES OF SOURCES IN THE SECOND MAXI/GSC CATALOG.

| (1) | (2) | (3) | (4) | (5) | (6) | (7) | (8) | (9) | (10) | (11) | (12) | (13) | (14) | (15) | (16) | (17) |
|---|---|---|---|---|---|---|---|---|---|---|---|---|---|---|---|---|
| | MAXI | | | | | | | | | Counterpart | | | | | | |
| No. | Name | R.A. | Decl. | $\sigma_{\rm stat}$[a] | $s_{\rm D,4-10keV}$ | $f_{4-10keV}$[b] | $s_{\rm D,3-4keV}$ | $f_{3-4keV}$[c] | HR | Name[d] | R.A. | Decl. | $z$ | Type[e] | Flag[f] | Note[g] |
| 1 | 2MAXI J0000−774 | 0.0 | −77.4 | 0.170 | 10.4 | 8.0±0.8 | 7.6 | 1.9±0.3 | 0.15±0.08 | | | | | | | |
| 2 | 2MAXI J0004+160 | 1.1 | 16.0 | 0.141 | 7.5 | 10.0±1.3 | 4.6 | 1.7±0.4 | 0.31±0.12 | RXC J0005.3+1612 | 1.344 | 16.211 | 0.1164 | Galaxy Cluster | M | |
| 3 | 2MAXI J0004+726 | 1.2 | 72.7 | 0.118 | 11.1 | 10.0±0.9 | 12.3 | 3.6±0.3 | −0.05±0.06 | | | | | | | |
| 4 | 2MAXI J0008−691 | 2.1 | −69.2 | 0.120 | 9.2 | 6.7±0.7 | 5.8 | 1.5±0.3 | 0.20±0.10 | | | | | | | |
| 5 | 2MAXI J0009−817 | 2.5 | −81.8 | 0.250 | 9.8 | 5.4±0.5 | 6.1 | 1.6±0.3 | 0.05±0.10 | | | | | | | |
| 6 | 2MAXI J0011−112 | 2.8 | −11.2 | 0.122 | 9.0 | 9.7±1.1 | 6.5 | 2.2±0.3 | 0.18±0.09 | | | | | | | |
| 7 | 2MAXI J0012−152 | 3.1 | −15.2 | 0.184 | 7.6 | 7.5±1.0 | 7.5 | 2.5±0.3 | 0.00±0.09 | MACSJ0011.7−1523 | 2.928 | −15.389 | 0.378 | Galaxy Cluster | M | |
| 8 | 2MAXI J0012−294 | 3.1 | −29.5 | 0.148 | 17.0 | 9.1±0.5 | 10.0 | 3.5±0.4 | −0.09±0.06 | | | | | | | |
| 9 | 2MAXI J0021+286 | 5.4 | 28.7 | 0.132 | 8.3 | 11.1±1.3 | 9.1 | 3.4±0.4 | 0.03±0.08 | RXC J0020.6+2840 | 5.170 | 28.675 | 0.094 | Galaxy Cluster | M | |
| 10 | 2MAXI J0023−193 | 5.8 | −19.3 | 0.130 | 10.1 | 10.8±1.1 | 9.4 | 3.1±0.3 | 0.06±0.07 | | | | | | | |
| 11 | 2MAXI J0037−790 | 9.3 | −79.1 | 0.130 | 17.0 | 8.9±0.5 | 9.9 | 2.5±0.3 | 0.07±0.06 | 2MASX J00341665−7905204 | 8.570 | −79.089 | 0.074 | Sy1 | B | |
| 12 | 2MAXI J0041−285 | 10.3 | −28.6 | 0.119 | 7.6 | 8.2±1.1 | 4.4 | 1.5±0.3 | 0.28±0.12 | RXC J0042.1−2832 | 10.536 | −28.536 | 0.1082 | Galaxy Cluster | M | |
| 13 | 2MAXI J0041−092 | 10.4 | −9.2 | 0.030 | 43.9 | 53.3±1.2 | 41.5 | 16.5±0.4 | 0.03±0.02 | ABELL 85 (1) | 10.408 | −9.342 | 0.0551 | Galaxy Cluster | BMG | |
| 14 | 2MAXI J0043+412 | 10.8 | 41.3 | 0.045 | 26.6 | 38.8±1.5 | 27.8 | 12.2±0.4 | 0.02±0.03 | M31 (2) | 10.668 | 41.200 | | Galaxy | BG | (A) |
| 15 | 2MAXI J0043+247 | 10.9 | 24.7 | 0.205 | 7.5 | 8.3±1.1 | 8.8 | 3.2±0.4 | −0.07±0.09 | ZwCl235 | 10.967 | 24.402 | 0.083 | Galaxy Cluster | M | |
| 16 | 2MAXI J0044−238 | 11.0 | −23.8 | 0.209 | 7.3 | 7.8±1.1 | 1.7 | 0.6±0.3 | 0.64±0.18 | NGC 235A | 10.720 | −23.541 | 0.0222 | Sy1 | B | |
| 17 | 2MAXI J0047−754 | 11.8 | −75.4 | 0.159 | 8.5 | 6.6±0.8 | 8.4 | 2.1±0.3 | 0.01±0.08 | | | | | | | |
| 18 | 2MAXI J0048+320 | 12.2 | 32.0 | 0.077 | 16.0 | 21.4±1.3 | 7.3 | 2.8±0.4 | 0.42±0.04 | Mrk 348 (3) | 12.196 | 31.957 | 0.015 | Sy2/FSRQ | BG | |
| 19 | 2MAXI J0051−725 | 12.8 | −72.6 | 0.038 | 29.6 | 24.1±0.8 | 21.1 | 5.7±0.3 | 0.16±0.03 | 2E0050.1−7247 (4) | 12.972 | −72.530 | | HMXB | G | |
| 20 | 2MAXI J0052+176 | 13.0 | 17.6 | 0.136 | 10.1 | 12.2±1.2 | 7.7 | 2.9±0.4 | 0.16±0.08 | Mrk 1148 | 12.978 | 17.433 | 0.064 | Sy1 | B | |
| 21 | 2MAXI J0055+463 | 13.9 | 46.3 | 0.120 | 9.8 | 13.1±1.3 | 7.0 | 2.7±0.4 | 0.24±0.08 | 1RXS J005528.0+461143 (5) | 13.833 | 46.216 | | CV/DQ Her | BG | |
| 22 | 2MAXI J0055+259 | 14.0 | 26.0 | 0.173 | 7.9 | 9.4±1.2 | 9.1 | 3.3±0.4 | −0.04±0.08 | A115 | 13.999 | 26.378 | 0.1971 | Galaxy Cluster | M | |
| 23 | 2MAXI J0056−011 | 14.1 | −1.2 | 0.060 | 21.5 | 26.1±1.2 | 23.8 | 9.4±0.4 | −0.04±0.03 | RXC J0056.3−0112 (6) | 14.076 | −1.217 | 0.0442 | Galaxy Cluster | MG | |
| 24 | 2MAXI J0102−222 | 15.5 | −22.3 | 0.140 | 9.5 | 10.5±1.1 | 13.6 | 4.7±0.3 | −0.15±0.06 | | | | | | | |
| 25 | 2MAXI J0105−241 | 16.4 | −24.2 | 0.216 | 7.6 | 7.6±1.0 | 7.0 | 2.5±0.4 | 0.01±0.10 | RXC J0104.5−2400 (C) | 16.146 | −24.000 | 0.1603 | Galaxy Cluster | M | (B) |
| 26 | 2MAXI J0107−468 | 16.8 | −46.8 | 0.248 | 10.3 | 12.0±1.2 | 5.2 | 2.0±0.4 | 0.32±0.10 | ESO 243− G 026 | 16.408 | −47.072 | 0.0193 | Galaxy | B | |
| 27 | 2MAXI J0111−148 | 18.0 | −14.9 | 0.388 | 7.7 | 7.9±1.0 | 9.8 | 3.3±0.3 | −0.12±0.08 | Mrk 1152 | 18.459 | −14.846 | 0.0527 | Sy1.5 | BM | |
| 28 | 2MAXI J0117−734 | 19.3 | −73.4 | 0.004 | 285.0 | 318.5±1.1 | 139.5 | 54.9±0.4 | 0.31±0.00 | SMC X-1 (7) | 19.271 | −73.443 | | HMXB/NS | BG | |
| 29 | 2MAXI J0123+342 | 20.8 | 34.3 | 0.127 | 11.3 | 15.2±1.3 | 14.5 | 5.8±0.4 | −0.08±0.06 | SHBL J012308.7+342049 (8) | 20.786 | 34.347 | 0.272 | BL Lac | BG | |
| 30 | 2MAXI J0124−588 | 21.0 | −58.8 | 0.065 | 18.6 | 17.9±1.0 | 15.0 | 4.3±0.3 | 0.15±0.04 | Fairall 9 | 20.941 | −58.806 | 0.047 | Sy1 | B | |
| 31 | 2MAXI J0124−349 | 21.0 | −35.0 | 0.069 | 18.6 | 22.5±1.2 | 14.3 | 5.4±0.4 | 0.16±0.04 | NGC 526A | 20.977 | −35.065 | 0.0191 | Sy1.5 | B | |
| 32 | 2MAXI J0126+081 | 21.5 | 8.2 | 0.177 | 8.4 | 10.3±1.2 | 9.3 | 3.5±0.4 | −0.02±0.08 | | | | | | | |
| 33 | 2MAXI J0134−082 | 23.6 | −8.2 | 0.174 | 7.1 | 7.9±1.1 | 5.9 | 2.0±0.3 | 0.12±0.11 | RXC J0132.6−0804 | 23.170 | −8.072 | 0.1489 | Galaxy Cluster | M | |
| 34 | 2MAXI J0151+361 | 28.0 | 36.1 | 0.087 | 11.2 | 14.9±1.3 | 17.5 | 7.7±0.4 | −0.22±0.05 | RXC J0152.7+3609 | 28.195 | 36.151 | 0.0163 | Galaxy Cluster | M | |
| 35 | 2MAXI J0157−838 | 29.3 | −83.9 | 0.188 | 7.7 | 5.9±0.8 | 5.9 | 1.6±0.3 | 0.09±0.11 | | | | | | | |
| 36 | 2MAXI J0157−528 | 29.4 | −52.8 | 0.091 | 9.2 | 11.1±1.2 | 9.5 | 3.0±0.3 | 0.09±0.08 | | | | | | | |
| 37 | 2MAXI J0202−727 | 30.7 | −72.8 | 0.128 | 10.9 | 8.3±0.8 | 4.4 | 1.1±0.2 | 0.43±0.10 | | | | | | | |
| 38 | 2MAXI J0205−020 | 31.4 | −2.1 | 0.147 | 9.6 | 11.3±1.2 | 7.5 | 2.7±0.4 | 0.16±0.08 | | | | | | | |
| 39 | 2MAXI J0206−166 | 31.6 | −16.6 | 0.179 | 8.7 | 9.2±1.1 | 6.7 | 2.2±0.3 | 0.16±0.09 | | | | | | | |
| 40 | 2MAXI J0206+153 | 31.7 | 15.3 | 0.130 | 9.8 | 13.1±1.3 | 6.3 | 2.5±0.4 | 0.26±0.09 | RX J0206.3+1511 | 31.598 | 15.188 | 0.248 | Galaxy Cluster | M | |
| 41 | 2MAXI J0207+751 | 31.8 | 75.1 | 0.283 | 8.2 | 7.3±0.9 | 3.2 | 1.0±0.3 | 0.41±0.14 | | | | | | | |
| 42 | 2MAXI J0213−637 | 33.4 | −63.8 | 0.223 | 10.0 | 6.9±0.7 | 5.5 | 1.4±0.3 | 0.22±0.10 | | | | | | | |
| 43 | 2MAXI J0224+186 | 36.1 | 18.6 | 0.080 | 12.0 | 14.3±1.2 | 9.6 | 3.5±0.4 | 0.14±0.07 | | | | | | | |
| 44 | 2MAXI J0228+314 | 37.1 | 31.4 | 0.103 | 12.3 | 18.0±1.5 | 10.9 | 4.3±0.4 | 0.15±0.06 | NGC 931 (9) | 37.060 | 31.312 | 0.0167 | Sy1.5 | BG | |
| 45 | 2MAXI J0231−440 | 37.8 | −44.1 | 0.171 | 8.6 | 10.5±1.2 | 9.4 | 3.7±0.4 | −0.04±0.08 | RXC J0232.2−4420 | 38.070 | −44.347 | 0.2836 | Galaxy Cluster | M | |
| 46 | 2MAXI J0232−090 | 38.1 | −9.1 | 0.167 | 7.9 | 9.2±1.2 | 8.7 | 3.1±0.4 | −0.01±0.09 | | | | | | | |
| 47 | 2MAXI J0233+324 | 38.4 | 32.5 | 0.106 | 11.4 | 16.6±1.5 | 9.8 | 3.9±0.4 | 0.17±0.07 | NGC 973 | 38.584 | 32.506 | 0.0162 | Sy2 | B | |

TABLE 1 — CONTINUED

| (1) | (2) | (3) | (4) | (5) | (6) | (7) | (8) | (9) | (10) | (11) | (12) | (13) | (14) | (15) | (16) | (17) |
|---|---|---|---|---|---|---|---|---|---|---|---|---|---|---|---|---|
| | MAXI | | | | | | | | | Counterpart | | | | | | |
| No. | Name | R.A. | Decl. | $\sigma_{stat}$[a] | $s_{D,4-10keV}$ | $f_{4-10keV}$[b] | $s_{D,3-4keV}$ | $f_{3-4keV}$[c] | HR | Name[d] | R.A. | Decl. | z | Type[e] | Flag[f] | Note[g] |
| 48 | 2MAXI J0236+025 | 39.1 | 2.5 | 0.142 | 7.7 | 8.1±1.1 | 4.9 | 1.8±0.4 | 0.20±0.12 | | | | | | | |
| 49 | 2MAXI J0238−308 | 39.7 | −30.8 | 0.191 | 9.1 | 9.6±1.1 | 6.3 | 2.2±0.4 | 0.17±0.09 | | | | | | | |
| 50 | 2MAXI J0239−522 | 39.9 | −52.3 | 0.070 | 14.1 | 13.5±1.0 | 11.4 | 3.7±0.3 | 0.09±0.06 | | | | | | | |
| 51 | 2MAXI J0243−582 | 40.8 | −58.2 | 0.125 | 12.7 | 11.1±0.9 | 11.4 | 3.2±0.3 | 0.06±0.06 | | | | | | | |
| 52 | 2MAXI J0251+468 | 42.9 | 46.9 | 0.155 | 7.7 | 8.9±1.2 | 5.3 | 2.0±0.4 | 0.19±0.11 | 2MASX J02502722+4647295 | 42.613 | 46.791 | | Galaxy | B | |
| 53 | 2MAXI J0254+416 | 43.6 | 41.7 | 0.028 | 43.6 | 63.5±1.5 | 51.7 | 26.7±0.5 | −0.12±0.01 | RXC J0254.4+4134 (10) | 43.623 | 41.572 | 0.0172 | Galaxy Cluster | MG | |
| 54 | 2MAXI J0257+197 | 44.3 | 19.7 | 0.180 | 8.1 | 9.5±1.2 | 4.6 | 1.7±0.4 | 0.30±0.11 | XY Ari | 44.038 | 19.441 | | CV/DQ Her | B | |
| 55 | 2MAXI J0258+135 | 44.6 | 13.5 | 0.029 | 43.3 | 63.1±1.5 | 40.9 | 19.5±0.5 | 0.03±0.02 | ABELL 401 (11) | 44.737 | 13.582 | 0.0748 | Galaxy Cluster | BMG | |
| 56 | 2MAXI J0259+060 | 44.8 | 6.0 | 0.152 | 8.2 | 9.7±1.2 | 8.7 | 3.3±0.4 | −0.02±0.08 | A400 | 44.412 | 6.006 | 0.0238 | Galaxy Cluster | M | |
| 57 | 2MAXI J0300+444 | 45.1 | 44.5 | 0.052 | 24.7 | 33.0±1.3 | 26.6 | 11.7±0.4 | −0.04±0.03 | RXC J0300.7+4427 (12) | 45.188 | 44.463 | 0.03 | Galaxy Cluster | MG | |
| 58 | 2MAXI J0300−107 | 45.2 | −10.7 | 0.098 | 15.5 | 14.9±1.0 | 8.8 | 3.0±0.3 | 0.24±0.06 | MCG −02−08−038 | 45.018 | −10.825 | 0.0326 | Sy1 | B | |
| 59 | 2MAXI J0303−726 | 46.0 | −72.7 | 0.140 | 7.6 | 4.7±0.6 | 6.9 | 1.7±0.2 | −0.05±0.10 | ESO 031− G 008 | 46.897 | −72.834 | 0.0276 | Sy1.2 | B | |
| 60 | 2MAXI J0306−233 | 46.6 | −23.4 | 0.189 | 7.4 | 8.2±1.1 | 3.1 | 1.1±0.3 | 0.44±0.14 | | | | | | | |
| 61 | 2MAXI J0307−115 | 47.0 | −11.6 | 0.344 | 7.1 | 7.7±1.1 | 5.3 | 1.8±0.3 | 0.17±0.11 | | | | | | | |
| 62 | 2MAXI J0308+410 | 47.1 | 41.0 | 0.050 | 25.8 | 37.6±1.5 | 33.0 | 17.1±0.5 | −0.16±0.02 | Algol (13) | 47.042 | 40.958 | | AlgolTypeEclipsingbinary | G | |
| 63 | 2MAXI J0313−773 | 48.3 | −77.4 | 0.184 | 8.2 | 6.4±0.8 | 6.1 | 1.5±0.2 | 0.16±0.10 | | | | | | | |
| 64 | 2MAXI J0317−443 | 49.4 | −44.3 | 0.085 | 15.1 | 20.1±1.3 | 16.2 | 7.1±0.4 | −0.04±0.05 | RXC J0317.9−4414 (14) | 49.494 | −44.239 | 0.0752 | Galaxy Cluster | MG | |
| 65 | 2MAXI J0319+415 | 49.9 | 41.5 | 0.004 | 276.2 | 637.6±2.3 | 202.8 | 201.7±1.0 | 0.02±0.00 | Perseus Cluster (15) | 49.951 | 41.512 | 0.0176 | Galaxy Cluster | BMG | |
| 66 | 2MAXI J0326+287 | 51.5 | 28.8 | 0.093 | 13.6 | 16.6±1.2 | 18.6 | 7.3±0.4 | −0.15±0.04 | UX Ari (16) | 51.647 | 28.715 | | RSCVn | G | |
| 67 | 2MAXI J0329−528 | 52.3 | −52.8 | 0.138 | 10.1 | 10.3±1.0 | 8.6 | 2.7±0.3 | 0.11±0.08 | RXC J0330.0−5235 | 52.503 | −52.596 | 0.0624 | Galaxy Cluster | M | |
| 68 | 2MAXI J0330+438 | 52.7 | 43.9 | 0.067 | 18.0 | 26.3±1.5 | 8.2 | 3.6±0.4 | 0.41±0.06 | GK Per (17) | 52.799 | 43.905 | | CV/DQ Her | BG | |
| 69 | 2MAXI J0334−360 | 53.5 | −36.1 | 0.132 | 11.2 | 12.9±1.2 | 9.6 | 3.6±0.4 | 0.07±0.07 | NGC 1365 (18) | 53.402 | −36.140 | 0.0055 | Sy1.8 | BG | |
| 70 | 2MAXI J0335+321 | 54.0 | 32.2 | 0.113 | 11.0 | 14.7±1.3 | 9.6 | 3.8±0.4 | 0.11±0.07 | 4C +32.14 | 54.126 | 32.308 | 1.258 | QSO | B | |
| 71 | 2MAXI J0336+006 | 54.1 | 0.6 | 0.082 | 16.3 | 19.8±1.2 | 25.9 | 10.3±0.4 | −0.23±0.03 | HR 1099 (19) | 54.197 | 0.588 | | RSCVn | G | |
| 72 | 2MAXI J0339+100 | 54.8 | 10.1 | 0.050 | 25.5 | 31.0±1.2 | 29.1 | 12.7±0.4 | −0.11±0.03 | RXC J0338.6+0958 (20) | 54.670 | 9.974 | 0.0347 | Galaxy Cluster | MG | |
| 73 | 2MAXI J0342−214 | 55.5 | −21.4 | 0.102 | 13.5 | 14.6±1.1 | 11.8 | 3.8±0.3 | 0.11±0.06 | ESO 548−G081 | 55.516 | −21.244 | 0.0145 | Sy1 | B | |
| 74 | 2MAXI J0343−536 | 55.9 | −53.7 | 0.052 | 23.1 | 24.7±1.1 | 23.8 | 7.9±0.3 | 0.01±0.03 | RXC J0342.8−5338 (21) | 55.725 | −53.635 | 0.059 | Galaxy Cluster | MG | |
| 75 | 2MAXI J0343+676 | 55.9 | 67.7 | 0.206 | 7.3 | 6.4±0.9 | 5.3 | 1.5±0.3 | 0.15±0.11 | | | | | | | |
| 76 | 2MAXI J0345+008 | 56.3 | 0.8 | 0.159 | 8.5 | 10.3±1.2 | 4.1 | 1.5±0.4 | 0.39±0.12 | | | | | | | |
| 77 | 2MAXI J0348−120 | 57.2 | −12.0 | 0.141 | 8.5 | 9.0±1.1 | 4.9 | 1.6±0.3 | 0.30±0.11 | QSO B0347−121 | 57.347 | −11.991 | 0.18 | BL Lac | BM | |
| 78 | 2MAXI J0354−738 | 58.6 | −73.9 | 0.155 | 13.8 | 9.2±0.7 | 12.9 | 3.3±0.3 | −0.04±0.05 | MS0353.3−7411 | 58.123 | −74.031 | 0.127 | Galaxy Cluster | M | |
| 79 | 2MAXI J0354−402 | 58.7 | −40.3 | 0.184 | 7.1 | 8.7±1.2 | 3.9 | 1.5±0.4 | 0.32±0.13 | | | | | | | |
| 80 | 2MAXI J0355+311 | 58.9 | 31.1 | 0.004 | 274.7 | 567.3±2.1 | 168.1 | 133.7±0.8 | 0.16±0.00 | X Per (22) | 58.846 | 31.046 | | HMXB/NS | BG | |
| 81 | 2MAXI J0358−819 | 59.7 | −81.9 | 0.122 | 8.9 | 5.9±0.7 | 7.2 | 1.9±0.3 | 0.02±0.09 | | | | | | | |
| 82 | 2MAXI J0405+381 | 61.5 | 38.2 | 0.087 | 11.4 | 15.2±1.3 | 12.9 | 5.6±0.4 | −0.06±0.06 | | | | | | | |
| 83 | 2MAXI J0408−644 | 62.2 | −64.4 | 0.185 | 8.7 | 7.3±0.8 | 7.5 | 2.0±0.3 | 0.09±0.09 | | | | | | | |
| 84 | 2MAXI J0413+106 | 63.3 | 10.6 | 0.029 | 38.5 | 51.4±1.3 | 33.1 | 14.5±0.4 | 0.08±0.02 | | | | | | | |
| 85 | 2MAXI J0423−752 | 66.0 | −75.3 | 0.227 | 9.6 | 6.2±0.6 | 6.7 | 1.7±0.3 | 0.09±0.09 | | | | | | | |
| 86 | 2MAXI J0424−571 | 66.2 | −57.1 | 0.107 | 12.2 | 13.4±1.1 | 9.0 | 2.6±0.3 | 0.25±0.06 | 1H 0419−577 (24) | 66.504 | −57.200 | 0.104 | Sy1.5 | BG | |
| 87 | 2MAXI J0425−198 | 66.4 | −19.8 | 0.152 | 9.4 | 9.7±1.0 | 7.7 | 2.5±0.3 | 0.13±0.08 | IW Eri | 66.480 | −19.758 | | CV/AM HER | B | |
| 88 | 2MAXI J0425−086 | 66.4 | −8.7 | 0.104 | 14.1 | 13.7±1.0 | 13.4 | 4.8±0.4 | −0.03±0.05 | RXC J0425.8−0833 | 66.464 | −8.559 | 0.0397 | Galaxy Cluster | M | |
| 89 | 2MAXI J0431−613 | 68.0 | −61.4 | 0.026 | 48.6 | 45.4±0.9 | 42.5 | 13.5±0.3 | 0.05±0.02 | ABELL 3266 (25) | 67.800 | −61.406 | 0.0589 | Galaxy Cluster | BMG | |
| 90 | 2MAXI J0433+054 | 68.4 | 5.5 | 0.064 | 19.0 | 25.4±1.3 | 20.3 | 8.1±0.4 | 0.01±0.04 | 3C 120 (26) | 68.296 | 5.354 | 0.033 | Sy1 | BG | |
| 91 | 2MAXI J0433−131 | 68.5 | −13.2 | 0.039 | 38.9 | 39.7±1.0 | 37.8 | 14.6±0.4 | −0.06±0.02 | RXC J0433.6−1315 | 68.410 | −13.259 | 0.0326 | Galaxy Cluster | M | |
| 92 | 2MAXI J0437+206 | 69.3 | 20.6 | 0.156 | 7.6 | 6.5±0.9 | 8.1 | 3.0±0.4 | −0.17±0.09 | | | | | | | |
| 93 | 2MAXI J0437−106 | 69.4 | −10.7 | 0.113 | 10.9 | 12.5±1.2 | 9.4 | 3.2±0.3 | 0.12±0.07 | MCG −02−12−050 | 69.559 | −10.796 | 0.0364 | Sy1.2 | B | |
| 94 | 2MAXI J0439−468 | 70.0 | −46.9 | 0.389 | 7.6 | 8.1±1.1 | 6.4 | 2.5±0.4 | 0.02±0.10 | 2MASX J04372814−4711298 | 69.367 | −47.191 | 0.053 | Sy1 | B | |
| 95 | 2MAXI J0440−048 | 70.1 | −4.8 | 0.175 | 10.3 | 8.6±0.8 | 6.4 | 2.3±0.4 | 0.11±0.09 | | | | | | | |

TABLE 1 — CONTINUED

| (1) | (2) | (3) | (4) | (5) | (6) | (7) | (8) | (9) | (10) | (11) | (12) | (13) | (14) | (15) | (16) | (17) |
|---|---|---|---|---|---|---|---|---|---|---|---|---|---|---|---|---|
| | MAXI | | | | | | | | | Counterpart | | | | | | |
| No. | Name | R.A. | Decl. | $\sigma_{stat}$[a] | $s_{D,4-10keV}$ | $f_{4-10keV}$[b] | $s_{D,3-4keV}$ | $f_{3-4keV}$[c] | HR | Name[d] | R.A. | Decl. | $z$ | Type[e] | Flag[f] | Note[g] |
| 96 | 2MAXI J0442−271 | 70.5 | −27.2 | 0.262 | 9.2 | 10.1±1.1 | 6.8 | 2.4±0.3 | 0.17±0.09 | IRAS 04392−2713 | 70.344 | −27.139 | 0.0835 | Sy1.5 | B | |
| 97 | 2MAXI J0443+288 | 70.8 | 28.8 | 0.131 | 9.7 | 11.8±1.2 | 4.7 | 1.7±0.4 | 0.38±0.10 | UGC 03142 | 70.945 | 28.972 | 0.0217 | Sy1 | B | |
| 98 | 2MAXI J0447−207 | 71.9 | −20.8 | 0.240 | 7.1 | 7.3±1.0 | 7.6 | 2.5±0.3 | −0.02±0.10 | RXC J0448.2−2028 | 72.051 | −20.470 | 0.072 | Galaxy Cluster | M | |
| 99 | 2MAXI J0449−094 | 72.4 | −9.4 | 0.225 | 7.3 | 6.8±0.9 | 3.2 | 1.1±0.4 | 0.32±0.15 | | | | | | | |
| 100 | 2MAXI J0450−035 | 72.6 | −3.5 | 0.137 | 10.9 | 13.2±1.2 | 6.8 | 2.5±0.4 | 0.28±0.08 | | | | | | | |
| 101 | 2MAXI J0451−585 | 72.9 | −58.6 | 0.141 | 8.8 | 8.4±0.9 | 4.5 | 1.2±0.3 | 0.38±0.11 | | | | | | | |
| 102 | 2MAXI J0453−752 | 73.4 | −75.3 | 0.123 | 19.0 | 13.4±0.7 | 15.6 | 4.1±0.3 | 0.03±0.04 | ESO 033− G 002 | 73.996 | −75.541 | 0.0181 | Sy2 | B | |
| 103 | 2MAXI J0453+070 | 73.4 | 7.0 | 0.148 | 8.3 | 10.0±1.2 | 25.8 | 10.3±0.4 | −0.51±0.05 | | | | | | | |
| 104 | 2MAXI J0457−696 | 74.4 | −69.6 | 0.075 | 19.0 | 15.0±0.8 | 9.8 | 2.6±0.3 | 0.31±0.05 | | | | | | | |
| 105 | 2MAXI J0502+247 | 75.6 | 24.7 | 0.113 | 13.6 | 16.2±1.2 | 8.1 | 3.1±0.4 | 0.27±0.07 | V1062 Tau | 75.615 | 24.756 | | Nova | B | |
| 106 | 2MAXI J0504−239 | 76.2 | −24.0 | 0.163 | 8.0 | 8.8±1.1 | 3.1 | 1.0±0.3 | 0.47±0.13 | 2MASX J05054575−2351139 | 76.441 | −23.854 | 0.035 | Sy2 HII | B | |
| 107 | 2MAXI J0508−044 | 77.2 | −4.4 | 0.209 | 7.1 | 7.6±1.1 | 3.5 | 1.2±0.4 | 0.34±0.14 | | | | | | | |
| 108 | 2MAXI J0509+677 | 77.5 | 67.8 | 0.104 | 12.4 | 10.8±0.9 | 11.6 | 3.5±0.3 | 0.01±0.06 | 87GB 050246.4+673341 | 76.984 | 67.623 | 0.314 | QSO | B | |
| 109 | 2MAXI J0510+167 | 77.6 | 16.7 | 0.064 | 18.1 | 28.7±1.6 | 13.6 | 5.4±0.4 | 0.27±0.04 | | | | | | | |
| 110 | 2MAXI J0514−399 | 78.6 | −40.0 | 0.022 | 53.6 | 78.2±1.5 | 51.5 | 24.6±0.5 | 0.02±0.01 | NGC1851 (30) | 78.527 | −40.044 | | LMXB/NS in globular c | BG | (D) |
| 111 | 2MAXI J0515−616 | 78.9 | −61.6 | 0.190 | 8.6 | 6.8±0.8 | 6.7 | 1.8±0.3 | 0.10±0.09 | | | | | | | |
| 112 | 2MAXI J0516−001 | 79.1 | −0.1 | 0.064 | 18.4 | 20.8±1.1 | 16.9 | 6.3±0.4 | 0.04±0.04 | Ark 120 (31) | 79.048 | −0.150 | 0.0323 | Sy1 | BG | |
| 113 | 2MAXI J0516−455 | 79.2 | −45.6 | 0.151 | 9.4 | 12.6±1.3 | 7.6 | 3.0±0.4 | 0.16±0.08 | | | | | | | |
| 114 | 2MAXI J0516−102 | 79.2 | −10.2 | 0.184 | 8.4 | 8.7±1.0 | 6.7 | 2.2±0.3 | 0.13±0.09 | MCG −02−14−009 | 79.088 | −10.562 | 0.0285 | Sy1 | B | |
| 115 | 2MAXI J0516+172 | 79.2 | 17.2 | 0.150 | 8.7 | 13.7±1.6 | 8.0 | 3.2±0.4 | 0.17±0.08 | RXC J0516.3+1712 | 79.086 | 17.209 | 0.115 | Galaxy Cluster | M | |
| 116 | 2MAXI J0517+065 | 79.4 | 6.5 | 0.141 | 9.2 | 11.0±1.2 | 12.5 | 4.8±0.4 | −0.14±0.07 | RXC J0516.6+0626 | 79.155 | 6.438 | 0.0284 | Galaxy Cluster | M | |
| 117 | 2MAXI J0518−325 | 79.6 | −32.5 | 0.123 | 8.3 | 9.4±1.1 | 7.4 | 2.7±0.4 | 0.07±0.09 | ESO 362-18 | 79.899 | −32.658 | 0.0125 | Sy1.5 | B | |
| 118 | 2MAXI J0520−719 | 80.1 | −71.9 | 0.005 | 241.0 | 278.1±1.2 | 191.3 | 91.3±0.5 | −0.00±0.00 | LMC X-2 (32) | 80.117 | −71.965 | | LMXB | BG | |
| 119 | 2MAXI J0522+631 | 80.6 | 63.2 | 0.147 | 8.8 | 8.7±1.0 | 7.9 | 2.4±0.3 | 0.08±0.08 | | | | | | | |
| 120 | 2MAXI J0523−363 | 80.8 | −36.3 | 0.117 | 10.8 | 12.7±1.2 | 9.6 | 3.5±0.4 | 0.08±0.07 | PKS 0521−36 | 80.742 | −36.459 | 0.0553 | BL Lac | BM | |
| 121 | 2MAXI J0529−327 | 82.3 | −32.7 | 0.039 | 34.6 | 42.1±1.2 | 17.7 | 6.7±0.4 | 0.35±0.03 | TV Col (33) | 82.356 | −32.818 | | CV/DQ Her | BG | |
| 122 | 2MAXI J0529−114 | 82.5 | −11.5 | 0.143 | 7.6 | 7.7±1.0 | 9.1 | 3.0±0.3 | −0.10±0.09 | | | | | | | |
| 123 | 2MAXI J0532−663 | 83.2 | −66.4 | 0.013 | 93.8 | 91.2±1.0 | 51.2 | 16.7±0.3 | 0.28±0.01 | LMC X-4 (34) | 83.207 | −66.371 | | HMXB/NS | BG | |
| 124 | 2MAXI J0533−024 | 83.4 | −2.5 | 0.207 | 8.6 | 10.4±1.2 | 8.9 | 3.3±0.4 | 0.01±0.08 | | | | | | | |
| 125 | 2MAXI J0535−580 | 83.8 | −58.1 | 0.067 | 16.1 | 14.9±0.9 | 10.0 | 2.9±0.3 | 0.26±0.05 | TW Pic | 83.711 | −58.028 | | CV DQ Her | BM | |
| 126 | 2MAXI J0535−052 | 83.8 | −5.3 | 0.027 | 46.1 | 61.7±1.3 | 54.9 | 26.2±0.5 | −0.13±0.01 | Trapezium Cluster (35) | 83.819 | −5.387 | | Star Cluster | BG | |
| 127 | 2MAXI J0539−640 | 84.8 | −64.1 | 0.007 | 146.4 | 170.7±1.2 | 194.4 | 100.5±0.5 | −0.29±0.00 | LMC X-3 (36) | 84.735 | −64.084 | | HMXB/NS | BG | |
| 128 | 2MAXI J0539+696 | 85.0 | −69.7 | 0.007 | 162.6 | 165.9±1.0 | 206.1 | 106.6±0.5 | −0.32±0.00 | LMC X-1 (37) | 84.911 | −69.743 | | HMXB/NS | BG | |
| 129 | 2MAXI J0541−019 | 85.3 | −2.0 | 0.098 | 13.9 | 16.9±1.2 | 16.1 | 6.2±0.4 | −0.06±0.05 | | | | | | | |
| 130 | 2MAXI J0541−089 | 85.3 | −8.9 | 0.172 | 7.3 | 7.2±1.0 | 2.2 | 0.7±0.3 | 0.52±0.17 | | | | | | | |
| 131 | 2MAXI J0542−407 | 85.6 | −40.8 | 0.104 | 11.7 | 14.2±1.2 | 8.4 | 3.2±0.4 | 0.19±0.07 | BVH2007 73 | 85.712 | −41.001 | 0.642 | Galaxy Cluster | M | |
| 132 | 2MAXI J0542+609 | 85.6 | 60.9 | 0.074 | 16.6 | 20.2±1.2 | 11.5 | 3.8±0.3 | 0.27±0.05 | BY Cam (38) | 85.704 | 60.859 | | CV/AM Her | BG | |
| 133 | 2MAXI J0547+594 | 86.8 | 59.5 | 0.177 | 7.4 | 9.0±1.2 | 9.8 | 3.2±0.3 | −0.05±0.08 | | | | | | | |
| 134 | 2MAXI J0549−621 | 87.5 | −62.2 | 0.137 | 9.5 | 7.6±0.8 | 8.5 | 2.5±0.3 | 0.00±0.08 | | | | | | | |
| 135 | 2MAXI J0550−320 | 87.6 | −32.0 | 0.071 | 13.9 | 21.9±1.6 | 17.3 | 6.9±0.4 | 0.02±0.05 | | | | | | | |
| 136 | 2MAXI J0552−073 | 88.1 | −7.4 | 0.021 | 53.2 | 71.1±1.3 | 33.5 | 13.3±0.4 | 0.27±0.02 | NGC 2110 (40) | 88.047 | −7.456 | 0.0078 | Sy2 | BG | |
| 137 | 2MAXI J0552−208 | 88.2 | −20.8 | 0.184 | 8.0 | 8.2±1.0 | 9.9 | 3.3±0.3 | −0.10±0.08 | RXC J0552.8−2103 | 88.218 | −21.057 | 0.0989 | Galaxy Cluster | M | |
| 138 | 2MAXI J0553−819 | 88.3 | −81.9 | 0.189 | 8.0 | 6.6±0.8 | 3.6 | 0.9±0.3 | 0.40±0.13 | | | | | | | |
| 139 | 2MAXI J0554+464 | 88.7 | 46.5 | 0.044 | 25.2 | 33.7±1.3 | 23.2 | 9.2±0.4 | 0.09±0.03 | MCG +08−11−011 (41) | 88.723 | 46.439 | 0.0205 | Sy1.5 | BG | |
| 140 | 2MAXI J0556−331 | 89.2 | −33.1 | 0.014 | 82.7 | 140.7±1.7 | 83.3 | 46.4±0.6 | −0.00±0.01 | | | | | | | |
| 141 | 2MAXI J0558+540 | 89.6 | 54.1 | 0.070 | 13.9 | 16.9±1.2 | 7.6 | 2.7±0.4 | 0.34±0.07 | V405 Aur | 89.497 | 53.896 | | CV/DQ Her | B | |
| 142 | 2MAXI J0558−499 | 89.7 | −49.9 | 0.156 | 7.1 | 7.8±1.1 | 6.5 | 2.2±0.3 | 0.09±0.10 | | | | | | | |
| 143 | 2MAXI J0601−397 | 90.3 | −39.7 | 0.131 | 8.8 | 10.7±1.2 | 11.0 | 4.3±0.4 | −0.09±0.07 | RXC J0601.7−3959 | 90.440 | −39.993 | 0.0468 | Galaxy Cluster | M | |

TABLE 1 — CONTINUED

| (1) | (2) | (3) | (4) | (5) | (6) | (7) | (8) | (9) | (10) | (11) | (12) | (13) | (14) | (15) | (16) | (17) |
|---|---|---|---|---|---|---|---|---|---|---|---|---|---|---|---|---|
| | MAXI | | | | | | | | | Counterpart | | | | | | |
| No. | Name | R.A. | Decl. | $\sigma_{stat}$[a] | $s_{D,4-10keV}$ | $f_{4-10keV}$[b] | $s_{D,3-4keV}$ | $f_{3-4keV}$[c] | HR | Name[d] | R.A. | Decl. | z | Type[e] | Flag[f] | Note[g] |
| 144 | 2MAXI J0607−352 | 91.9 | −35.3 | 0.125 | 8.8 | 10.5±1.2 | 5.2 | 1.9±0.4 | 0.29±0.10 | RXC J0605.8−3518 | 91.470 | −35.301 | 0.1392 | Galaxy Cluster | M | |
| 145 | 2MAXI J0613−656 | 93.3 | −65.7 | 0.177 | 8.9 | 6.4±0.7 | 10.2 | 2.7±0.3 | −0.13±0.07 | | | | | | | |
| 146 | 2MAXI J0614−217 | 93.7 | −21.8 | 0.136 | 8.9 | 9.5±1.1 | 7.5 | 2.4±0.3 | 0.12±0.09 | | | | | | | |
| 147 | 2MAXI J0615+712 | 94.0 | 71.3 | 0.167 | 8.2 | 7.2±0.9 | 1.6 | 0.4±0.3 | 0.69±0.17 | Mrk 3 | 93.901 | 71.037 | 0.0135 | Sy2 | M | |
| 148 | 2MAXI J0623−095 | 95.9 | −9.6 | 0.156 | 7.9 | 8.4±1.1 | 2.2 | 0.7±0.3 | 0.59±0.16 | SWIFT J062406.05−093855.0 | 93.901 | −9.649 | | SRC/X-RAY | B | |
| 149 | 2MAXI J0627−543 | 96.9 | −54.3 | 0.075 | 10.5 | 25.5±2.4 | 25.9 | 10.0±0.4 | −0.09±0.05 | RXC J0627.2−5428 (42) | 96.810 | −54.470 | 0.0506 | Galaxy Cluster | M | |
| 150 | 2MAXI J0632+738 | 98.2 | 73.9 | 0.163 | 12.3 | 10.9±0.9 | 6.9 | 2.0±0.3 | 0.29±0.08 | | | | | | | |
| 151 | 2MAXI J0634−751 | 98.7 | −75.2 | 0.142 | 8.3 | 6.8±0.8 | 7.6 | 1.9±0.2 | 0.08±0.09 | PKS 0637−752 | 98.944 | −75.271 | 0.651 | Sy1 | B | |
| 152 | 2MAXI J0639+279 | 100.0 | 27.9 | 0.254 | 7.6 | 5.6±0.7 | 2.2 | 0.8±0.4 | 0.39±0.20 | | | | | | | |
| 153 | 2MAXI J0640−257 | 100.2 | −25.7 | 0.216 | 7.6 | 7.8±1.0 | 7.8 | 2.6±0.3 | −0.01±0.09 | ESO 490−IG026 | 100.049 | −25.895 | 0.0248 | Sy1.2 | B | |
| 154 | 2MAXI J0641−540 | 100.4 | −54.0 | 0.123 | 9.4 | 9.7±1.0 | 10.1 | 3.1±0.3 | 0.02±0.07 | | | | | | | |
| 155 | 2MAXI J0644−554 | 101.2 | 55.4 | 0.115 | 12.0 | 13.9±1.2 | 6.3 | 2.2±0.4 | 0.34±0.08 | | | | | | | |
| 156 | 2MAXI J0656−560 | 104.2 | −56.1 | 0.093 | 11.5 | 12.2±1.1 | 9.6 | 2.8±0.3 | 0.17±0.07 | Bullet Cluster | 104.622 | −55.953 | 0.296 | Galaxy Cluster | BM | |
| 157 | 2MAXI J0706−573 | 106.7 | −57.3 | 0.195 | 7.4 | 4.8±0.6 | 3.8 | 1.1±0.3 | 0.18±0.14 | | | | | | | |
| 158 | 2MAXI J0707+641 | 106.9 | 64.1 | 0.205 | 7.1 | 6.6±0.9 | 7.5 | 2.2±0.3 | −0.01±0.10 | VII Zw 118 | 106.805 | 64.600 | 0.0797 | Sy1.0 | B | |
| 159 | 2MAXI J0710+592 | 107.6 | 59.2 | 0.086 | 14.8 | 15.4±1.0 | 15.4 | 5.1±0.3 | −0.00±0.05 | 2MASX J07103005+5908202 (43) | 107.625 | 59.139 | 0.125 | BL Lac | BG | |
| 160 | 2MAXI J0715−363 | 108.8 | −36.4 | 0.298 | 8.1 | 9.7±1.2 | 11.5 | 4.3±0.4 | −0.15±0.07 | RXC J0717.1−3621 | 109.295 | −36.360 | 0.032 | Galaxy Cluster | M | |
| 161 | 2MAXI J0715+442 | 109.0 | 44.3 | 0.113 | 11.1 | 14.8±1.3 | 9.8 | 3.9±0.4 | 0.11±0.07 | | | | | | | |
| 162 | 2MAXI J0717−494 | 109.3 | −49.5 | 0.233 | 7.1 | 7.4±1.0 | 4.2 | 1.5±0.4 | 0.25±0.13 | | | | | | | |
| 163 | 2MAXI J0721+555 | 110.4 | 55.6 | 0.149 | 9.6 | 10.8±1.1 | 12.3 | 4.5±0.4 | −0.12±0.07 | RXC J0721.3+5547 | 110.343 | 55.786 | 0.0381 | Galaxy Cluster | M | |
| 164 | 2MAXI J0723−731 | 110.9 | −73.2 | 0.146 | 9.7 | 7.5±0.8 | 6.6 | 1.6±0.3 | 0.20±0.09 | | | | | | | |
| 165 | 2MAXI J0730+100 | 112.7 | 10.0 | 0.111 | 10.5 | 14.0±1.3 | 4.2 | 1.7±0.4 | 0.47±0.10 | BG CMi | 112.871 | 9.940 | | CV/DQ Her | B | |
| 166 | 2MAXI J0740+497 | 115.2 | 49.8 | 0.158 | 7.3 | 8.9±1.2 | 7.4 | 2.8±0.4 | 0.03±0.10 | Mrk 79 | 115.637 | 49.810 | 0.0222 | Sy1.2 | B | |
| 167 | 2MAXI J0742−547 | 115.5 | −54.7 | 0.205 | 9.1 | 8.7±1.0 | 7.5 | 2.2±0.3 | 0.12±0.09 | 2MASX J07410919−5447461 | 115.288 | −54.796 | | Galaxy | B | |
| 168 | 2MAXI J0742+745 | 115.6 | 74.5 | 0.226 | 7.1 | 6.0±0.8 | 7.2 | 2.0±0.3 | −0.02±0.10 | RXC J0741.7+7414 | 115.437 | 74.248 | 0.2149 | Galaxy Cluster | M | |
| 169 | 2MAXI J0744+288 | 116.1 | 28.8 | 0.152 | 7.4 | 8.2±1.1 | 7.5 | 2.7±0.4 | −0.01±0.09 | | | | | | | |
| 170 | 2MAXI J0745−530 | 116.3 | −53.1 | 0.213 | 9.8 | 9.4±1.0 | 7.6 | 2.3±0.3 | 0.14±0.08 | V0436 Car (44) | 116.243 | −52.952 | | CV/DQHer | G | |
| 171 | 2MAXI J0751+149 | 117.8 | 15.0 | 0.091 | 12.4 | 21.2±1.7 | 6.3 | 3.0±0.5 | 0.40±0.08 | PQ Gem | 117.823 | 14.740 | | CV/DQ Her | B | |
| 172 | 2MAXI J0751+182 | 117.9 | 18.2 | 0.171 | 8.4 | 9.6±1.2 | 6.9 | 2.6±0.4 | 0.10±0.09 | | | | | | | |
| 173 | 2MAXI J0752+221 | 118.2 | 22.2 | 0.353 | 8.1 | 6.2±0.8 | 4.6 | 1.6±0.4 | 0.11±0.12 | | | | | | | |
| 174 | 2MAXI J0758+632 | 119.6 | 63.2 | 0.248 | 7.2 | 6.9±1.0 | 7.3 | 2.3±0.3 | 0.01±0.10 | | | | | | | |
| 175 | 2MAXI J0801+162 | 120.3 | 16.2 | 0.141 | 8.0 | 10.6±1.3 | 7.3 | 2.8±0.4 | 0.11±0.09 | | | | | | | |
| 176 | 2MAXI J0804+052 | 121.0 | 5.2 | 0.144 | 8.0 | 9.7±1.2 | 2.5 | 0.9±0.4 | 0.55±0.14 | Mrk 1210 | 121.024 | 5.114 | 0.0135 | Sy2 | B | |
| 177 | 2MAXI J0808+757 | 122.2 | 75.7 | 0.134 | 9.0 | 8.5±0.9 | 10.4 | 3.0±0.3 | −0.03±0.07 | PG 0804+761 | 122.744 | 76.045 | 0.1 | Sy1 | B | |
| 178 | 2MAXI J0814+627 | 123.7 | 62.7 | 0.114 | 8.2 | 8.4±1.0 | 8.7 | 2.7±0.3 | 0.01±0.08 | MS0811.6+6301 | 123.999 | 62.886 | 0.312 | Galaxy Cluster | M | |
| 179 | 2MAXI J0815−572 | 124.0 | −57.2 | 0.079 | 14.5 | 14.1±1.0 | 18.3 | 5.5±0.3 | −0.09±0.04 | | | | | | | |
| 180 | 2MAXI J0817−074 | 124.4 | −7.4 | 0.026 | 31.2 | 32.5±1.0 | 26.4 | 9.9±0.4 | 0.04±0.02 | RXC J0817.4−0730 (46) | 124.352 | −7.513 | 0.0704 | Galaxy Cluster | MG | |
| 181 | 2MAXI J0817+017 | 124.4 | 1.7 | 0.094 | 9.4 | 10.7±1.1 | 8.0 | 2.9±0.4 | 0.09±0.08 | | | | | | | |
| 182 | 2MAXI J0823+082 | 125.8 | 8.2 | 0.237 | 7.5 | 6.9±0.9 | 4.6 | 1.7±0.4 | 0.15±0.12 | | | | | | | |
| 183 | 2MAXI J0825+733 | 126.3 | 73.3 | 0.103 | 12.5 | 11.2±0.9 | 7.4 | 2.1±0.3 | 0.28±0.07 | | | | | | | |
| 184 | 2MAXI J0827−642 | 126.8 | −64.3 | 0.187 | 7.8 | 6.2±0.8 | 6.7 | 1.8±0.3 | 0.07±0.10 | | | | | | | |
| 185 | 2MAXI J0831+660 | 127.8 | 66.0 | 0.173 | 7.2 | 6.4±0.9 | 9.3 | 2.7±0.3 | −0.13±0.09 | RXC J0830.9+6551 | 127.743 | 65.850 | 0.1818 | Galaxy Cluster | M | |
| 186 | 2MAXI J0831−701 | 127.9 | −70.1 | 0.154 | 9.9 | 7.3±0.7 | 2.5 | 0.6±0.2 | 0.59±0.13 | | | | | | | |
| 187 | 2MAXI J0835−040 | 129.0 | −4.1 | 0.159 | 8.0 | 9.4±1.2 | 6.2 | 2.2±0.4 | 0.17±0.10 | | | | | | | |
| 188 | 2MAXI J0839+485 | 129.8 | 48.5 | 0.122 | 10.3 | 12.5±1.2 | 6.4 | 2.4±0.4 | 0.25±0.09 | EI UMa | 129.592 | 48.634 | | CV/DQ Her | B | |
| 189 | 2MAXI J0840−125 | 130.0 | −12.6 | 0.151 | 9.7 | 9.6±1.0 | 6.9 | 2.2±0.3 | 0.17±0.09 | 3C 206 | 129.961 | −12.243 | 0.1976 | QSO | B | |
| 190 | 2MAXI J0840+708 | 130.2 | 70.8 | 0.100 | 11.9 | 10.8±0.9 | 6.3 | 1.8±0.3 | 0.33±0.08 | [HB89] 0836+710 | 130.351 | 70.895 | 2.172 | Blazar | BM | |
| 191 | 2MAXI J0846+142 | 131.7 | 14.2 | 0.233 | 7.4 | 8.0±1.1 | 4.7 | 1.8±0.4 | 0.19±0.12 | 2MASX J08451850+1420345 | 131.327 | 14.343 | 0.0606 | Sy1.9 | B | |

TABLE 1 — CONTINUED

| (1) | (2) | (3) | (4) | (5) | (6) | (7) | (8) | (9) | (10) | (11) | (12) | (13) | (14) | (15) | (16) | (17) |
|---|---|---|---|---|---|---|---|---|---|---|---|---|---|---|---|---|
| | MAXI | | | | | | | | | Counterpart | | | | | | |
| No. | Name | R.A. | Decl. | $\sigma_{stat}$[a] | $s_{D,4-10keV}$ | $f_{4-10keV}$[b] | $s_{D,3-4keV}$ | $f_{3-4keV}$[c] | HR | Name[d] | R.A. | Decl. | z | Type[e] | Flag[f] | Note[g] |
| 192 | 2MAXI J0854−074 | 133.6 | −7.5 | 0.214 | 7.8 | 7.9±1.0 | 6.2 | 2.1±0.3 | 0.09±0.10 | | | | | | | |
| 193 | 2MAXI J0856−247 | 134.0 | −24.8 | 0.145 | 10.4 | 11.1±1.1 | 8.4 | 2.8±0.3 | 0.12±0.08 | | | | | | | |
| 194 | 2MAXI J0909−096 | 137.3 | −9.6 | 0.021 | 58.6 | 71.2±1.2 | 49.2 | 19.6±0.4 | 0.09±0.01 | ABELL 754 (47) | 137.209 | −9.637 | 0.0542 | Galaxy Cluster | BMG | |
| 195 | 2MAXI J0910+524 | 137.7 | 52.5 | 0.194 | 9.5 | 11.5±1.2 | 5.1 | 1.8±0.4 | 0.35±0.10 | | | | | | | |
| 196 | 2MAXI J0918−119 | 139.5 | −12.0 | 0.072 | 17.0 | 19.0±1.1 | 21.4 | 7.4±0.3 | −0.09±0.04 | RXC J0918.1−1205 (48) | 139.527 | −12.093 | 0.0539 | Galaxy Cluster | MG | |
| 197 | 2MAXI J0920−079 | 140.2 | −7.9 | 0.092 | 14.7 | 17.0±1.2 | 10.3 | 3.6±0.4 | 0.21±0.06 | MCG −01−24−012 | 140.193 | −8.056 | 0.0196 | Sy2 | B | |
| 198 | 2MAXI J0921−224 | 140.4 | −22.4 | 0.230 | 8.3 | 9.2±1.1 | 8.3 | 2.7±0.3 | 0.05±0.09 | | | | | | | |
| 199 | 2MAXI J0922+301 | 140.6 | 30.1 | 0.239 | 7.1 | 8.6±1.2 | 5.2 | 1.9±0.4 | 0.18±0.12 | RXC J0920.4+3030 | 140.109 | 30.515 | 0.2952 | Galaxy Cluster | M | |
| 200 | 2MAXI J0923+228 | 140.8 | 22.9 | 0.130 | 9.2 | 10.3±1.1 | 6.5 | 2.3±0.4 | 0.20±0.09 | MCG +04−22−042 | 140.929 | 22.909 | 0.0323 | Sy1.2 | B | |
| 201 | 2MAXI J0923+521 | 141.0 | 52.1 | 0.074 | 17.0 | 20.7±1.2 | 15.2 | 5.7±0.4 | 0.09±0.04 | | | | | | | |
| 202 | 2MAXI J0924−316 | 141.0 | −31.6 | 0.025 | 47.5 | 63.5±1.3 | 46.2 | 20.2±0.4 | 0.01±0.02 | 2MASX J09235371−3141305 (49) | 140.974 | −31.692 | 0.0422 | Sy1.9 | BG | |
| 203 | 2MAXI J0945−140 | 146.5 | −14.1 | 0.145 | 9.5 | 9.2±1.0 | 4.2 | 1.3±0.3 | 0.39±0.11 | NGC 2992 | 146.425 | −14.326 | 0.0077 | Sy2 | BM | |
| 204 | 2MAXI J0947+288 | 146.9 | 28.9 | 0.191 | 7.1 | 8.2±1.2 | 4.2 | 1.5±0.4 | 0.28±0.13 | | | | | | | |
| 205 | 2MAXI J0947−308 | 146.9 | −30.9 | 0.020 | 62.0 | 82.9±1.3 | 43.2 | 18.9±0.4 | 0.18±0.01 | MCG −05−23−016 (50) | 146.917 | −30.949 | 0.0085 | Sy2 | BG | |
| 206 | 2MAXI J0949+757 | 147.3 | 75.7 | 0.255 | 7.3 | 6.5±0.9 | 6.3 | 1.8±0.3 | 0.09±0.10 | | | | | | | |
| 207 | 2MAXI J0953−763 | 148.4 | −76.3 | 0.161 | 9.8 | 7.7±0.8 | 7.4 | 1.9±0.3 | 0.15±0.08 | | | | | | | |
| 208 | 2MAXI J0955−694 | 148.8 | −69.4 | 0.282 | 9.1 | 7.3±0.8 | 4.9 | 1.3±0.3 | 0.31±0.11 | | | | | | | |
| 209 | 2MAXI J0956+694 | 149.2 | 69.4 | 0.048 | 27.6 | 24.2±0.9 | 22.9 | 7.1±0.3 | 0.05±0.03 | MAXI J0957+693 (51) | 149.300 | 69.400 | | confused | | (E) |
| 210 | 2MAXI J1000−313 | 150.2 | −31.3 | 0.091 | 13.3 | 14.3±1.1 | 8.6 | 3.1±0.4 | 0.19±0.07 | 2MASX J09594263−3112581 | 149.928 | −31.216 | 0.037 | Sy1 | B | |
| 211 | 2MAXI J1007+662 | 151.8 | 66.3 | 0.198 | 8.3 | 6.5±0.8 | 4.2 | 1.2±0.3 | 0.27±0.12 | | | | | | | |
| 212 | 2MAXI J1009−311 | 152.4 | −31.2 | 0.098 | 13.9 | 16.8±1.2 | 13.6 | 5.0±0.4 | 0.05±0.05 | | | | | | | |
| 213 | 2MAXI J1016−274 | 154.1 | −27.5 | 0.213 | 8.7 | 9.4±1.1 | 6.7 | 2.4±0.4 | 0.13±0.09 | | | | | | | |
| 214 | 2MAXI J1020−034 | 155.1 | −3.4 | 0.165 | 8.2 | 9.6±1.2 | 9.0 | 3.2±0.4 | −0.00±0.08 | ARK 241 | 155.418 | −3.454 | 0.0408 | Sy1 | B | |
| 215 | 2MAXI J1023+199 | 155.9 | 19.9 | 0.064 | 19.5 | 23.2±1.2 | 15.8 | 5.8±0.4 | 0.14±0.04 | NGC 3227 (52) | 155.877 | 19.865 | 0.0039 | Sy1.5 | BG | |
| 216 | 2MAXI J1025+499 | 156.3 | 49.9 | 0.250 | 7.1 | 8.6±1.2 | 8.5 | 3.1±0.4 | −0.05±0.09 | RXC J1022.5+5006 | 155.627 | 50.102 | 0.158 | Galaxy Cluster | M | |
| 217 | 2MAXI J1028+729 | 157.2 | 73.0 | 0.202 | 7.5 | 7.0±0.9 | 3.4 | 0.9±0.3 | 0.41±0.13 | CGCG 333−038 | 158.598 | 73.014 | 0.022 | Galaxy | B | |
| 218 | 2MAXI J1030−351 | 157.7 | −35.2 | 0.140 | 8.9 | 10.8±1.2 | 12.4 | 4.5±0.4 | −0.12±0.07 | | | | | | | |
| 219 | 2MAXI J1031−143 | 158.0 | −14.3 | 0.109 | 11.1 | 11.7±1.1 | 7.6 | 2.4±0.3 | 0.22±0.08 | 2MASSi J1031543−141651 | 157.976 | −14.281 | 0.086 | Sy1 | B | |
| 220 | 2MAXI J1036−275 | 159.1 | −27.6 | 0.046 | 26.5 | 32.1±1.2 | 36.8 | 14.6±0.4 | −0.16±0.02 | RXC J1036.6−2731 (53) | 159.174 | −27.524 | 0.0126 | Galaxy Cluster | MG | |
| 221 | 2MAXI J1037−104 | 159.4 | −10.5 | 0.223 | 7.1 | 6.8±1.0 | 7.5 | 2.5±0.3 | −0.05±0.10 | | | | | | | |
| 222 | 2MAXI J1103−234 | 165.8 | −23.4 | 0.105 | 12.3 | 13.8±1.1 | 17.8 | 6.2±0.3 | −0.15±0.05 | 2MASX J11033765−2329307 (54) | 165.907 | −23.492 | 0.186 | BL Lac | BG | |
| 223 | 2MAXI J1104+382 | 166.1 | 38.2 | 0.011 | 107.1 | 182.2±1.7 | 111.4 | 70.9±0.6 | −0.09±0.01 | Mrk 421 (55) | 166.114 | 38.209 | 0.03 | BL Lac | BG | |
| 224 | 2MAXI J1105+724 | 166.4 | 72.4 | 0.048 | 24.6 | 24.8±1.0 | 13.8 | 4.1±0.3 | 0.33±0.04 | NGC 3516 (56) | 166.698 | 72.569 | 0.0088 | Sy1.5 | BG | |
| 225 | 2MAXI J1108−056 | 167.1 | −5.6 | 0.216 | 7.2 | 7.9±1.1 | 5.0 | 1.8±0.4 | 0.19±0.12 | | | | | | | |
| 226 | 2MAXI J1116−768 | 169.2 | −76.8 | 0.173 | 9.3 | 7.2±0.8 | 12.7 | 3.2±0.3 | −0.16±0.07 | | | | | | | |
| 227 | 2MAXI J1123+120 | 170.9 | 12.0 | 0.325 | 8.4 | 6.7±0.8 | 5.9 | 2.2±0.4 | 0.00±0.10 | | | | | | | |
| 228 | 2MAXI J1127+190 | 171.8 | 19.0 | 0.209 | 7.9 | 7.3±0.9 | 4.9 | 1.7±0.4 | 0.17±0.12 | 2MASX J11271632+1909198 | 171.818 | 19.156 | 0.1055 | Sy1.8 | B | |
| 229 | 2MAXI J1129−144 | 172.3 | −14.5 | 0.096 | 11.6 | 12.6±1.1 | 9.1 | 3.0±0.3 | 0.16±0.07 | RXC J1130.3−1434 | 172.581 | −14.583 | 0.1068 | Galaxy Cluster | M | |
| 230 | 2MAXI J1132+146 | 173.1 | 14.6 | 0.138 | 8.8 | 10.4±1.2 | 8.3 | 3.2±0.4 | 0.04±0.08 | RXC J1132.8+1428 | 173.221 | 14.469 | 0.0834 | Galaxy Cluster | M | |
| 231 | 2MAXI J1136+675 | 174.0 | 67.6 | 0.090 | 15.2 | 14.4±0.9 | 16.9 | 5.1±0.3 | −0.04±0.04 | 2MASX J11363009+6737042 (57) | 174.125 | 67.618 | 0.1342 | BL Lac | BG | |
| 232 | 2MAXI J1138−119 | 174.7 | −11.9 | 0.286 | 7.5 | 7.3±1.0 | 6.1 | 2.0±0.3 | 0.09±0.10 | | | | | | | |
| 233 | 2MAXI J1139+220 | 174.8 | 22.1 | 0.155 | 9.0 | 9.2±1.0 | 5.4 | 1.9±0.4 | 0.23±0.10 | | | | | | | |
| 234 | 2MAXI J1139−376 | 174.9 | −37.7 | 0.040 | 29.7 | 39.7±1.3 | 21.6 | 8.4±0.4 | 0.21±0.03 | NGC 3783 (58) | 174.757 | −37.739 | 0.0097 | Sy1 | BG | |
| 235 | 2MAXI J1143+592 | 175.9 | 59.2 | 0.187 | 8.3 | 9.9±1.2 | 7.7 | 2.4±0.3 | 0.14±0.09 | MCG +10−17−061 | 176.388 | 58.978 | 0.0099 | Galaxy | B | |
| 236 | 2MAXI J1144+717 | 176.0 | 71.7 | 0.059 | 20.4 | 20.1±1.0 | 14.1 | 4.2±0.3 | 0.22±0.04 | DO Dra (59) | 175.910 | 71.689 | | CV/DQ Her | BG | |
| 237 | 2MAXI J1144+198 | 176.2 | 19.8 | 0.052 | 24.5 | 28.9±1.2 | 31.1 | 12.4±0.4 | −0.13±0.03 | A1367 (60) | 176.152 | 19.759 | 0.0214 | Galaxy Cluster | MG | |
| 238 | 2MAXI J1145−182 | 176.4 | −18.2 | 0.068 | 15.0 | 16.2±1.1 | 12.7 | 4.1±0.3 | 0.13±0.05 | | | | | | | |
| 239 | 2MAXI J1149−077 | 177.4 | −7.8 | 0.192 | 7.1 | 7.6±1.1 | 6.4 | 2.2±0.4 | 0.05±0.11 | | | | | | | |

TABLE 1 — Continued

| (1) | (2) | (3) | (4) | (5) | (6) | (7) | (8) | (9) | (10) | (11) | (12) | (13) | (14) | (15) | (16) | (17) |
|---|---|---|---|---|---|---|---|---|---|---|---|---|---|---|---|---|
| | MAXI | | | | | | | | | Counterpart | | | | | | |
| No. | Name | R.A. | Decl. | $\sigma_{stat}$[a] | $s_{D,4-10keV}$ | $f_{4-10keV}$[b] | $s_{D,3-4keV}$ | $f_{3-4keV}$[c] | HR | Name[d] | R.A. | Decl. | z | Type[e] | Flag[f] | Note[g] |
| 240 | 2MAXI J1151−119 | 177.8 | −11.9 | 0.155 | 9.1 | 9.3±1.0 | 6.8 | 2.3±0.3 | 0.14±0.09 | | | | | | | |
| 241 | 2MAXI J1155+234 | 178.8 | 23.5 | 0.135 | 10.7 | 12.6±1.2 | 12.2 | 4.5±0.4 | −0.04±0.06 | RXC J1155.3−2324 | 178.827 | 23.407 | 0.1427 | Galaxy Cluster | M | |
| 242 | 2MAXI J1158+562 | 179.7 | 56.3 | 0.177 | 8.0 | 8.2±1.0 | 7.3 | 2.5±0.3 | 0.03±0.09 | RXC J1200.3+5613 | 180.092 | 56.230 | 0.065 | Galaxy Cluster | M | |
| 243 | 2MAXI J1159−307 | 179.9 | −30.8 | 0.192 | 7.0 | 8.5±1.2 | 8.5 | 2.7±0.3 | 0.01±0.09 | | | | | | | |
| 244 | 2MAXI J1203+445 | 180.8 | 44.6 | 0.121 | 15.5 | 18.8±1.2 | 13.1 | 5.2±0.4 | 0.08±0.05 | NGC 4051 | 180.790 | 44.531 | 0.0023 | Sy1.5 | B | |
| 245 | 2MAXI J1209+476 | 182.4 | 47.6 | 0.200 | 7.3 | 8.0±1.1 | 3.1 | 1.1±0.4 | 0.40±0.15 | | | | | | | |
| 246 | 2MAXI J1210+394 | 182.7 | 39.4 | 0.015 | 91.8 | 122.6±1.3 | 31.6 | 15.1±0.5 | 0.45±0.01 | NGC 4151 (62) | 182.636 | 39.406 | 0.0033 | Sy1.5 | BG | |
| 247 | 2MAXI J1217+073 | 184.4 | 7.3 | 0.124 | 9.3 | 11.3±1.2 | 6.1 | 2.3±0.4 | 0.24±0.09 | NGC 4235 | 184.291 | 7.192 | 0.008 | Sy1 | B | |
| 248 | 2MAXI J1219+036 | 184.8 | 3.7 | 0.102 | 13.8 | 16.7±1.2 | 14.9 | 5.7±0.4 | −0.02±0.05 | | | | | | | |
| 249 | 2MAXI J1220+300 | 185.0 | 30.1 | 0.046 | 25.7 | 34.4±1.3 | 25.3 | 10.1±0.4 | 0.06±0.03 | | | | | | | |
| 250 | 2MAXI J1220+752 | 185.2 | 75.3 | 0.132 | 10.8 | 7.7±0.7 | 7.0 | 2.0±0.3 | 0.12±0.08 | Mrk 205 | 185.433 | 75.311 | 0.0708 | BL Lac | BM | |
| 251 | 2MAXI J1227−091 | 186.9 | −9.1 | 0.197 | 8.3 | 9.2±1.1 | 5.1 | 1.7±0.3 | 0.27±0.11 | | | | | | | |
| 252 | 2MAXI J1228−486 | 187.1 | −48.6 | 0.094 | 14.2 | 17.2±1.2 | 9.8 | 3.7±0.4 | 0.21±0.06 | 1RXS J122758.8−485343 | 186.995 | −48.896 | | CV? | B | |
| 253 | 2MAXI J1229+021 | 187.3 | 2.2 | 0.021 | 61.4 | 82.1±1.3 | 44.3 | 19.4±0.4 | 0.16±0.01 | | | | | | | |
| 254 | 2MAXI J1230+124 | 187.6 | 12.5 | 0.010 | 124.3 | 196.4±1.6 | 145.2 | 104.0±0.7 | −0.24±0.00 | Virgo galaxy cluster (65) | 187.706 | 12.391 | 0.00436 | GalaxyCluster | G | |
| 255 | 2MAXI J1237−387 | 189.4 | −38.7 | 0.093 | 13.9 | 16.9±1.2 | 10.0 | 3.7±0.4 | 0.19±0.06 | V1025 Cen (66) | 189.569 | −38.713 | | CV/DQ HER | BG | |
| 256 | 2MAXI J1240−049 | 190.0 | −5.0 | 0.083 | 16.3 | 18.7±1.1 | 12.6 | 4.7±0.4 | 0.14±0.05 | | | | | | | |
| 257 | 2MAXI J1240−116 | 190.2 | −11.7 | 0.211 | 7.3 | 8.1±1.1 | 3.8 | 1.3±0.3 | 0.36±0.13 | WARP J1240.4−1147 | 190.104 | −11.793 | 0.193 | Galaxy Cluster | M | |
| 258 | 2MAXI J1249+412 | 192.3 | 41.2 | 0.015 | 64.1 | 93.4±1.5 | 70.2 | 36.3±0.5 | −0.09±0.01 | RXC J1248.7−4118 (68) | 192.200 | −41.308 | 0.0114 | Galaxy Cluster | MG | |
| 259 | 2MAXI J1252−291 | 193.2 | −29.2 | 0.021 | 57.9 | 84.5±1.5 | 53.9 | 23.6±0.4 | 0.08±0.01 | | | | | | | |
| 260 | 2MAXI J1253−127 | 193.4 | −12.7 | 0.233 | 7.5 | 7.9±1.0 | 5.1 | 1.7±0.3 | 0.21±0.11 | RXC J1255.7−1239 | 193.925 | −12.655 | 0.0585 | Galaxy Cluster | M | |
| 261 | 2MAXI J1254−063 | 193.5 | −6.4 | 0.230 | 8.7 | 9.2±1.1 | 5.2 | 1.8±0.4 | 0.24±0.11 | RXC J1254.0−0642 | 193.511 | −6.701 | 0.1962 | Galaxy Cluster | M | |
| 262 | 2MAXI J1256−305 | 194.2 | −30.6 | 0.103 | 10.9 | 14.5±1.3 | 17.5 | 6.7±0.4 | −0.17±0.05 | RXC J1257.2−3022 | 194.320 | −30.377 | 0.0554 | Galaxy Cluster | M | |
| 263 | 2MAXI J1257−173 | 194.3 | −17.4 | 0.056 | 20.4 | 22.8±1.1 | 23.1 | 8.0±0.3 | −0.03±0.03 | RXC J1257.1−1724 (70) | 194.290 | −17.400 | 0.0473 | Galaxy Cluster | MG | |
| 264 | 2MAXI J1258+016 | 194.6 | 1.6 | 0.167 | 7.3 | 8.0±1.1 | 6.7 | 2.4±0.4 | 0.05±0.10 | | | | | | | |
| 265 | 2MAXI J1259−043 | 194.8 | −4.4 | 0.068 | 21.0 | 23.5±1.1 | 18.0 | 6.7±0.4 | 0.07±0.04 | RXC J1259.3−0411 | 194.840 | −4.195 | 0.0845 | Galaxy Cluster | M | |
| 266 | 2MAXI J1259+279 | 194.9 | 28.0 | 0.007 | 161.1 | 254.4±1.6 | 121.5 | 77.4±0.6 | 0.04±0.01 | Coma Cluster (71) | 194.953 | 27.981 | 0.0231 | Galaxy Cluster | BMG | |
| 267 | 2MAXI J1259−017 | 194.9 | −1.7 | 0.089 | 14.7 | 17.3±1.2 | 16.7 | 6.2±0.4 | −0.04±0.05 | RXC J1258.6−0145 | 194.671 | −1.757 | 0.0845 | Galaxy Cluster | M | |
| 268 | 2MAXI J1304+192 | 196.2 | 19.3 | 0.173 | 7.4 | 8.5±1.2 | 8.7 | 3.1±0.4 | −0.05±0.09 | RXC J1303.7+1916 | 195.940 | 19.271 | 0.0643 | Galaxy Cluster | M | |
| 269 | 2MAXI J1307−400 | 196.9 | −40.0 | 0.129 | 10.0 | 12.2±1.2 | 2.7 | 1.0±0.4 | 0.60±0.12 | | | | | | | |
| 270 | 2MAXI J1311−014 | 198.0 | −1.5 | 0.103 | 13.1 | 15.4±1.2 | 13.1 | 4.8±0.4 | 0.03±0.05 | RXC J1311.4−0120 | 197.875 | −1.335 | 0.1832 | Galaxy Cluster | M | |
| 271 | 2MAXI J1325−429 | 201.4 | −43.0 | 0.007 | 164.7 | 300.2±1.8 | 66.0 | 34.1±0.5 | 0.48±0.01 | Cen A (72) | 201.365 | −43.019 | 0.0018 | Sy2 | BG | |
| 272 | 2MAXI J1327−272 | 201.8 | −27.2 | 0.074 | 17.4 | 20.0±1.2 | 18.0 | 6.4±0.4 | 0.01±0.04 | RXC J1326.9−2710 | 201.725 | −27.183 | 0.0458 | Galaxy Cluster | M | |
| 273 | 2MAXI J1329−315 | 202.4 | −31.5 | 0.025 | 52.1 | 63.3±1.2 | 51.8 | 22.7±0.4 | −0.04±0.01 | RXC J1329.7−3136 | 202.429 | −31.602 | 0.0488 | Galaxy Cluster | M | |
| 274 | 2MAXI J1332+593 | 203.0 | 59.3 | 0.184 | 8.1 | 8.2±1.0 | 10.1 | 3.2±0.3 | −0.09±0.08 | | | | | | | |
| 275 | 2MAXI J1335−341 | 203.9 | −34.2 | 0.044 | 30.7 | 36.2±1.2 | 21.9 | 8.7±0.4 | 0.15±0.03 | MCG −06−30−015 (75) | 203.974 | −34.296 | 0.0077 | Galaxy Cluster | BG | |
| 276 | 2MAXI J1338+046 | 204.6 | 4.6 | 0.058 | 18.6 | 22.4±1.2 | 9.7 | 3.5±0.4 | 0.36±0.05 | NGC 5252 (77) | 204.566 | 4.543 | 0.023 | Sy1.9 | BG | |
| 277 | 2MAXI J1347−328 | 206.9 | −32.8 | 0.014 | 65.6 | 87.7±1.3 | 59.7 | 26.1±0.4 | 0.05±0.01 | ABELL 3571 (78) | 206.871 | −32.866 | 0.0391 | Galaxy Cluster | BMG | |
| 278 | 2MAXI J1348+024 | 207.1 | 2.4 | 0.173 | 8.3 | 9.0±1.1 | 2.8 | 0.9±0.3 | 0.52±0.14 | SWIFT J1347.7+0212 | 206.852 | 2.312 | | Sy1.2 | B | |
| 279 | 2MAXI J1348−117 | 207.2 | −11.7 | 0.142 | 9.2 | 10.2±1.1 | 6.4 | 2.1±0.3 | 0.23±0.09 | RXC J1347.5−1144 | 206.875 | −11.749 | 0.4516 | Galaxy Cluster | M | |
| 280 | 2MAXI J1348+267 | 207.2 | 26.7 | 0.034 | 35.8 | 41.8±1.2 | 35.7 | 14.2±0.4 | −0.02±0.02 | RXC J1348.8+2635 (79) | 207.221 | 26.596 | 0.0622 | Galaxy Cluster | MG | |
| 281 | 2MAXI J1349−302 | 207.3 | −30.2 | 0.021 | 62.3 | 75.7±1.2 | 47.1 | 18.7±0.4 | 0.14±0.01 | IC 4329A (80) | 207.330 | −30.309 | 0.016 | Sy1.2 | BMG | |
| 282 | 2MAXI J1353−739 | 208.4 | −73.9 | 0.062 | 9.9 | 7.7±0.8 | 5.8 | 1.5±0.3 | 0.24±0.09 | | | | | | | |
| 283 | 2MAXI J1357+214 | 209.3 | 21.5 | 0.209 | 8.0 | 8.4±1.1 | 2.2 | 0.8±0.3 | 0.57±0.16 | | | | | | | |
| 284 | 2MAXI J1357−094 | 209.3 | −9.5 | 0.191 | 7.9 | 8.6±1.1 | 6.2 | 2.1±0.3 | 0.15±0.10 | | | | | | | |
| 285 | 2MAXI J1358−478 | 209.7 | −47.9 | 0.058 | 20.3 | 27.2±1.3 | 21.0 | 8.4±0.4 | 0.03±0.03 | RXC J1358.9−4750 | 209.737 | −47.839 | 0.074 | Galaxy Cluster | M | |
| 286 | 2MAXI J1400+032 | 210.1 | 3.2 | 0.119 | 10.6 | 11.4±1.1 | 9.7 | 3.4±0.4 | 0.04±0.07 | | | | | | | |
| 287 | 2MAXI J1406−300 | 211.7 | −30.0 | 0.180 | 7.9 | 8.6±1.1 | 4.6 | 1.5±0.3 | 0.29±0.12 | 2MASX J14080674−3023537 | 212.028 | −30.398 | 0.0235 | Sy1.4 | B | |

TABLE 1 — CONTINUED

| (1) | (2) | (3) | (4) | (5) | (6) | (7) | (8) | (9) | (10) | (11) | (12) | (13) | (14) | (15) | (16) | (17) |
|---|---|---|---|---|---|---|---|---|---|---|---|---|---|---|---|---|
| | MAXI | | | | | | | | | Counterpart | | | | | | |
| No. | Name | R.A. | Decl. | $\sigma_{stat}$[a] | $s_{D,4-10keV}$ | $f_{4-10keV}$[b] | $s_{D,3-4keV}$ | $f_{3-4keV}$[c] | HR | Name[d] | R.A. | Decl. | z | Type[e] | Flag[f] | Note[g] |
| 288 | 2MAXI J1408−508 | 212.0 | −50.9 | 0.058 | 20.7 | 22.3±1.1 | 18.0 | 6.3±0.4 | 0.07±0.04 | RXC J1407.8−5100 (82) | 211.969 | −51.009 | 0.0966 | Galaxy Cluster | MG | |
| 289 | 2MAXI J1409−428 | 212.5 | −42.8 | 0.124 | 9.9 | 13.3±1.3 | 9.2 | 3.5±0.4 | 0.11±0.07 | RXC J1410.4−4246 | 212.619 | −42.777 | 0.049 | Galaxy Cluster | M | |
| 290 | 2MAXI J1413−030 | 213.3 | −3.1 | 0.026 | 49.6 | 60.3±1.2 | 32.1 | 12.5±0.4 | 0.22±0.02 | NGC 5506 (83) | 213.312 | −3.208 | 0.0062 | Sy1.9 | BG | |
| 291 | 2MAXI J1417+253 | 214.4 | 25.3 | 0.063 | 19.8 | 28.8±1.5 | 20.6 | 8.1±0.4 | 0.08±0.03 | NGC 5548 (84) | 214.498 | 25.137 | 0.0172 | Sy1.5 | BG | |
| 292 | 2MAXI J1417−128 | 214.5 | −12.9 | 0.300 | 7.2 | 6.0±0.8 | 3.2 | 1.0±0.3 | 0.31±0.15 | | | | | | | |
| 293 | 2MAXI J1419−265 | 214.9 | −26.6 | 0.103 | 9.2 | 10.1±1.1 | 5.6 | 1.9±0.3 | 0.28±0.10 | ESO 511−G030 | 214.843 | −26.645 | 0.0224 | Sy1 | B | |
| 294 | 2MAXI J1420−779 | 215.1 | −78.0 | 0.071 | 8.1 | 6.8±0.8 | 4.4 | 0.9±0.2 | 0.41±0.11 | | | | | | | |
| 295 | 2MAXI J1420+483 | 215.2 | 48.3 | 0.186 | 8.9 | 10.8±1.2 | 8.9 | 3.3±0.4 | 0.04±0.08 | | | | | | | |
| 296 | 2MAXI J1420−485 | 215.2 | −48.6 | 0.195 | 9.7 | 10.7±1.1 | 7.7 | 2.8±0.4 | 0.11±0.08 | | | | | | | |
| 297 | 2MAXI J1422−380 | 215.7 | −38.1 | 0.193 | 9.3 | 11.2±1.2 | 11.8 | 4.4±0.4 | −0.09±0.07 | | | | | | | |
| 298 | 2MAXI J1423+379 | 215.9 | 37.9 | 0.163 | 9.4 | 12.6±1.3 | 11.5 | 4.6±0.4 | −0.05±0.07 | BVH2007 NS 10 | 216.269 | 37.971 | 0.163 | Galaxy Cluster | M | |
| 299 | 2MAXI J1423+250 | 216.0 | 25.0 | 0.179 | 7.7 | 11.2±1.5 | 5.8 | 2.1±0.4 | 0.27±0.10 | NGC 5610 | 216.095 | 24.614 | 0.0169 | Sy2 | B | |
| 300 | 2MAXI J1428−730 | 217.1 | −73.0 | 0.062 | 9.0 | 7.0±0.8 | 7.5 | 1.8±0.2 | 0.11±0.09 | | | | | | | |
| 301 | 2MAXI J1429+425 | 217.4 | 42.5 | 0.078 | 14.2 | 18.9±1.3 | 15.3 | 6.7±0.4 | −0.04±0.05 | 1ES 1426+428 (85) | 217.136 | 42.672 | 0.129 | BL Lac | BMG | |
| 302 | 2MAXI J1436−364 | 219.1 | −36.4 | 0.106 | 12.4 | 14.5±1.2 | 7.2 | 2.6±0.4 | 0.30±0.07 | | | | | | | |
| 303 | 2MAXI J1437+588 | 219.4 | 58.8 | 0.152 | 10.0 | 9.6±1.0 | 8.5 | 2.7±0.3 | 0.07±0.08 | Mrk 817 | 219.092 | 58.794 | 0.0314 | Sy1.5 | B | |
| 304 | 2MAXI J1441−386 | 220.3 | −38.6 | 0.198 | 7.1 | 8.4±1.2 | 5.2 | 1.9±0.4 | 0.18±0.12 | | | | | | | |
| 305 | 2MAXI J1446+152 | 221.6 | 15.3 | 0.176 | 9.8 | 11.9±1.2 | 7.8 | 2.9±0.4 | 0.14±0.08 | | | | | | | |
| 306 | 2MAXI J1454−243 | 223.6 | −24.3 | 0.143 | 9.0 | 9.8±1.1 | 10.7 | 3.6±0.3 | −0.06±0.07 | | | | | | | |
| 307 | 2MAXI J1455−381 | 223.9 | −38.1 | 0.156 | 9.1 | 9.8±1.1 | 7.1 | 2.5±0.4 | 0.11±0.09 | RXC J1456.2−3826 | 224.051 | −38.440 | 0.115 | Galaxy Cluster | M | |
| 308 | 2MAXI J1459+217 | 224.9 | 21.7 | 0.160 | 9.2 | 10.4±1.1 | 11.1 | 3.9±0.4 | −0.07±0.07 | A2009 | 225.085 | 21.362 | 0.153 | Galaxy Cluster | M | |
| 309 | 2MAXI J1503−421 | 225.8 | −42.1 | 0.099 | 12.4 | 15.0±1.2 | 21.2 | 8.4±0.4 | −0.26±0.04 | | | | | | | |
| 310 | 2MAXI J1503+108 | 225.8 | 10.9 | 0.103 | 12.8 | 15.6±1.2 | 9.6 | 3.5±0.4 | 0.19±0.06 | | | | | | | |
| 311 | 2MAXI J1503−027 | 226.0 | −2.7 | 0.090 | 12.4 | 14.5±1.2 | 11.7 | 4.1±0.4 | 0.07±0.06 | RXC J1504.1−0248 | 226.032 | −2.805 | 0.2153 | Galaxy Cluster | M | |
| 312 | 2MAXI J1510−815 | 227.7 | −81.5 | 0.119 | 9.0 | 7.4±0.8 | 3.4 | 0.9±0.3 | 0.47±0.12 | 2MASX J15144217−8123377 | 228.675 | −81.394 | 0.0684 | Sy1.2 | B | |
| 313 | 2MAXI J1511+059 | 227.8 | 5.9 | 0.025 | 48.3 | 64.5±1.3 | 42.8 | 18.7±0.4 | 0.06±0.02 | | | | | | | |
| 314 | 2MAXI J1512−213 | 228.0 | −21.4 | 0.127 | 9.7 | 10.3±1.1 | 7.7 | 2.5±0.3 | 0.14±0.08 | 2MASX J15115979−2119015 | 227.999 | −21.317 | 0.0446 | Sy1/NL | B | |
| 315 | 2MAXI J1513+423 | 228.4 | 42.4 | 0.219 | 7.4 | 8.9±1.2 | 5.0 | 2.0±0.4 | 0.19±0.12 | NGC 5899 | 228.763 | 42.050 | 0.0086 | Sy2 | B | |
| 316 | 2MAXI J1521+082 | 230.4 | 8.2 | 0.091 | 21.8 | 24.1±1.1 | 20.5 | 9.0±0.4 | −0.06±0.03 | | | | | | | |
| 317 | 2MAXI J1522+278 | 230.6 | 27.9 | 0.070 | 17.6 | 21.1±1.2 | 19.3 | 7.3±0.4 | −0.03±0.04 | RXC J1522.4−2742 (88) | 230.610 | 27.709 | 0.0723 | Galaxy Cluster | MG | |
| 318 | 2MAXI J1523+304 | 230.8 | 30.5 | 0.119 | 11.5 | 13.5±1.2 | 12.6 | 4.9±0.4 | −0.05±0.06 | | | | | | | |
| 319 | 2MAXI J1523−300 | 231.0 | −30.1 | 0.314 | 8.2 | 8.6±1.1 | 4.9 | 1.7±0.3 | 0.26±0.11 | | | | | | | |
| 320 | 2MAXI J1535+581 | 233.9 | 58.1 | 0.143 | 8.1 | 8.0±1.0 | 7.2 | 2.3±0.3 | 0.06±0.09 | Mrk 290 | 233.968 | 57.903 | 0.0296 | Sy1 | B | |
| 321 | 2MAXI J1539−031 | 234.8 | −3.2 | 0.127 | 11.6 | 13.6±1.2 | 8.8 | 3.1±0.3 | 0.18±0.07 | RXC J1540.1−0318 | 235.031 | −3.308 | 0.1533 | Galaxy Cluster | M | |
| 322 | 2MAXI J1539+219 | 234.9 | 22.0 | 0.159 | 8.3 | 8.9±1.1 | 8.3 | 2.9±0.4 | 0.01±0.09 | A2107 | 234.910 | 21.789 | 0.0411 | Galaxy Cluster | M | |
| 323 | 2MAXI J1545+361 | 236.5 | 36.2 | 0.209 | 7.5 | 9.1±1.2 | 8.4 | 3.3±0.4 | −0.05±0.09 | A2124 | 236.250 | 36.066 | 0.0654 | Galaxy Cluster | M | |
| 324 | 2MAXI J1546−834 | 236.6 | −83.4 | 0.159 | 7.6 | 6.1±0.8 | 10.0 | 2.7±0.3 | −0.16±0.08 | RXC J1539.5−8335 | 234.891 | −83.592 | 0.0728 | Galaxy Cluster | M | |
| 325 | 2MAXI J1548−136 | 237.1 | −13.6 | 0.092 | 13.3 | 13.9±1.0 | 8.0 | 2.6±0.3 | 0.27±0.07 | NGC 5995 | 237.104 | −13.758 | 0.0252 | Sy2 | B | |
| 326 | 2MAXI J1551−224 | 238.0 | −22.5 | 0.167 | 8.8 | 9.0±1.0 | 5.0 | 1.7±0.3 | 0.28±0.11 | | | | | | | |
| 327 | 2MAXI J1552+850 | 238.2 | 85.1 | 0.198 | 8.6 | 8.1±0.9 | 5.5 | 1.6±0.3 | 0.24±0.10 | LEDA 100168 | 241.849 | 85.030 | 0.183 | Sy1 | B | |
| 328 | 2MAXI J1555−793 | 238.8 | −79.3 | 0.105 | 8.9 | 7.0±0.8 | 8.4 | 2.1±0.3 | 0.04±0.08 | PKS 1549−79 | 239.245 | −79.234 | 0.1501 | Sy1 | B | |
| 329 | 2MAXI J1555+112 | 239.0 | 11.2 | 0.157 | 10.5 | 11.0±1.0 | 9.7 | 3.5±0.4 | 0.01±0.07 | | | | | | | |
| 330 | 2MAXI J1558+272 | 239.6 | 27.2 | 0.031 | 30.9 | 56.3±1.8 | 34.7 | 15.2±0.4 | 0.10±0.02 | ABELL 2142 (90) | 239.567 | 27.225 | 0.0909 | Galaxy Cluster | BMG | |
| 331 | 2MAXI J1559−138 | 239.9 | −13.8 | 0.115 | 11.4 | 11.7±1.0 | 9.8 | 3.3±0.3 | 0.07±0.07 | | | | | | | |
| 332 | 2MAXI J1600+256 | 240.2 | 25.7 | 0.133 | 9.7 | 12.9±1.3 | 1.9 | 0.7±0.4 | 0.71±0.13 | | | | | | | |
| 333 | 2MAXI J1601−758 | 240.4 | −75.9 | 0.038 | 18.4 | 15.9±0.9 | 16.3 | 4.2±0.3 | 0.10±0.04 | RXC J1601.7−7544 (92) | 240.445 | −75.746 | 0.153 | Galaxy Cluster | MG | |
| 334 | 2MAXI J1602+161 | 240.7 | 16.2 | 0.050 | 24.7 | 33.0±1.3 | 28.3 | 12.4±0.4 | −0.07±0.03 | | | | | | | |
| 335 | 2MAXI J1603−099 | 240.9 | −9.9 | 0.217 | 8.3 | 9.0±1.1 | 6.6 | 2.2±0.3 | 0.13±0.10 | | | | | | | |

TABLE 1 — CONTINUED

| (1) | (2) | (3) | (4) | (5) | (6) | (7) | (8) | (9) | (10) | (11) | (12) | (13) | (14) | (15) | (16) | (17) |
|---|---|---|---|---|---|---|---|---|---|---|---|---|---|---|---|---|
| | MAXI | | | | | | | | | Counterpart | | | | | | |
| No. | Name | R.A. | Decl. | $\sigma_{stat}$[a] | $s_{D,4-10keV}$ | $f_{4-10keV}$[b] | $s_{D,3-4keV}$ | $f_{3-4keV}$[c] | HR | Name[d] | R.A. | Decl. | z | Type[e] | Flag[f] | Note[g] |
| 336 | 2MAXI J1605−196 | 241.3 | −19.7 | 0.157 | 15.7 | 14.7±0.9 | 12.8 | 4.6±0.4 | 0.02±0.05 | | | | | | | |
| 337 | 2MAXI J1607−727 | 241.8 | −72.8 | 0.122 | 11.2 | 8.0±0.7 | 6.2 | 1.5±0.2 | 0.27±0.09 | 2MASX J16052330−7253565 | 241.347 | −72.899 | | Galaxy | B | |
| 338 | 2MAXI J1613+812 | 243.4 | 81.2 | 0.271 | 7.7 | 6.4±0.8 | 3.5 | 1.0±0.3 | 0.35±0.14 | CGCG 367−009 | 244.830 | 81.046 | 0.0274 | Sy2 | B | |
| 339 | 2MAXI J1614+660 | 243.6 | 66.0 | 0.165 | 9.1 | 7.5±0.8 | 8.2 | 2.4±0.3 | 0.01±0.08 | Mrk 876 | 243.488 | 65.719 | 0.129 | Sy1 | B | |
| 340 | 2MAXI J1615−060 | 243.9 | −6.0 | 0.060 | 23.0 | 28.0±1.2 | 20.2 | 7.6±0.4 | 0.10±0.03 | RXC J1615.7−0608 (93) | 243.945 | −6.146 | 0.203 | Galaxy Cluster | MG | |
| 341 | 2MAXI J1615−109 | 243.9 | −10.9 | 0.223 | 9.9 | 11.0±1.1 | 9.2 | 3.3±0.4 | 0.03±0.07 | | | | | | | |
| 342 | 2MAXI J1616+258 | 244.0 | 25.9 | 0.205 | 7.0 | 7.4±1.1 | 4.7 | 1.7±0.4 | 0.18±0.12 | | | | | | | |
| 343 | 2MAXI J1617+504 | 244.4 | 50.4 | 0.247 | 7.9 | 6.9±0.9 | 1.2 | 0.4±0.4 | 0.67±0.22 | | | | | | | |
| 344 | 2MAXI J1619−281 | 244.8 | −28.2 | 0.045 | 28.7 | 34.8±1.2 | 24.5 | 9.4±0.4 | 0.09±0.03 | 2MASS J16193334−2807397 (94) | 244.889 | −28.128 | | Symb/NS | BG | |
| 345 | 2MAXI J1619+066 | 244.8 | 6.6 | 0.169 | 8.5 | 9.0±1.1 | 5.1 | 1.8±0.4 | 0.25±0.11 | | | | | | | |
| 346 | 2MAXI J1619−155 | 245.0 | −15.6 | <0.001 | 7747.4 | 178839.2±23.1 | 4889.3 | 52508.2±10.7 | 0.05±0.00 | Sco X-1 (95) | 244.980 | −15.640 | | LMXB/NS | BG | |
| 347 | 2MAXI J1622−231 | 245.6 | −23.1 | 0.154 | 8.1 | 10.8±1.3 | 9.3 | 3.4±0.4 | 0.01±0.08 | | | | | | | |
| 348 | 2MAXI J1626−333 | 246.7 | −33.3 | 0.062 | 20.4 | 24.5±1.2 | 17.8 | 6.7±0.4 | 0.09±0.04 | RXC J1626.3−3329 (96) | 246.586 | −33.489 | 0.1098 | Galaxy Cluster | MG | |
| 349 | 2MAXI J1627−245 | 246.8 | −24.6 | 0.053 | 25.9 | 31.2±1.2 | 29.3 | 11.6±0.4 | −0.07±0.03 | MAXI J1627−243 (97) | 246.800 | −24.400 | | SRC/X-RAY | | (F) |
| 350 | 2MAXI J1628+396 | 247.2 | 39.6 | 0.036 | 33.3 | 48.5±1.5 | 39.4 | 18.8±0.5 | −0.08±0.02 | RXC J1628.6+3932 (98) | 247.158 | 39.549 | 0.0299 | Galaxy Cluster | MG | |
| 351 | 2MAXI J1632−674 | 248.1 | −67.5 | 0.006 | 220.0 | 259.3±1.2 | 131.6 | 52.3±0.4 | 0.24±0.00 | 4U 1626−67 (99) | 248.070 | −67.462 | | LMXB | BG | |
| 352 | 2MAXI J1632+057 | 248.2 | 5.8 | 0.065 | 18.0 | 21.2±1.2 | 17.6 | 6.5±0.4 | 0.03±0.04 | RXC J1632.7+0534 (100) | 248.194 | 5.571 | 0.1514 | Galaxy Cluster | MG | |
| 353 | 2MAXI J1633−756 | 248.4 | −75.6 | 0.061 | 19.0 | 15.9±0.8 | 16.4 | 4.4±0.3 | 0.09±0.04 | | | | | | | |
| 354 | 2MAXI J1638−643 | 249.6 | −64.4 | 0.011 | 102.9 | 105.1±1.0 | 83.6 | 30.3±0.4 | 0.06±0.01 | TrA Cluster (102) | 249.567 | −64.347 | 0.0508 | Galaxy Cluster | BMG | |
| 355 | 2MAXI J1639+468 | 249.8 | 46.9 | 0.114 | 10.6 | 12.9±1.2 | 9.4 | 3.6±0.4 | 0.09±0.07 | RXC J1640.3+4642 | 250.089 | 46.706 | 0.228 | Galaxy Cluster | M | |
| 356 | 2MAXI J1646−735 | 251.5 | −73.5 | 0.129 | 11.8 | 8.8±0.7 | 15.4 | 4.0±0.3 | −0.16±0.05 | RXC J1645.4−7334 | 251.359 | −73.582 | 0.069 | Galaxy Cluster | M | |
| 357 | 2MAXI J1647−229 | 251.9 | −22.9 | 0.105 | 11.8 | 14.2±1.2 | 10.4 | 3.6±0.4 | 0.12±0.06 | | | | | | | |
| 358 | 2MAXI J1653+398 | 253.5 | 39.8 | 0.032 | 39.3 | 57.2±1.5 | 36.4 | 17.4±0.5 | 0.04±0.02 | Mrk 501 (103) | 253.468 | 39.760 | 0.0337 | BL Lac | BG | |
| 359 | 2MAXI J1658+353 | 254.5 | 35.4 | 0.007 | 174.4 | 317.8±1.8 | 79.1 | 47.2±0.6 | 0.38±0.01 | Her X-1 (104) | 254.458 | 35.342 | | LMXB/NS | BG | |
| 360 | 2MAXI J1659−151 | 254.8 | −15.1 | 0.016 | 69.1 | 83.9±1.2 | 77.9 | 34.1±0.4 | −0.11±0.01 | MAXI J1659−152 | 254.756 | −15.258 | | XRB/BH | B | |
| 361 | 2MAXI J1704+785 | 256.0 | 78.6 | 0.028 | 46.5 | 42.4±0.9 | 39.3 | 13.0±0.3 | 0.03±0.02 | ABELL 2256 (106) | 255.931 | 78.718 | 0.0581 | Galaxy Cluster | BMG | |
| 362 | 2MAXI J1704−011 | 256.2 | −1.2 | 0.093 | 14.4 | 16.6±1.2 | 13.4 | 4.9±0.4 | 0.05±0.05 | | | | | | | |
| 363 | 2MAXI J1705−617 | 256.4 | −61.8 | 0.077 | 15.9 | 14.7±0.9 | 19.7 | 5.9±0.3 | −0.10±0.04 | SWIFT J179616.4−614240 | 256.500 | −61.717 | 0.0809 | transient | B | |
| 364 | 2MAXI J1706+240 | 256.6 | 24.1 | 0.053 | 22.3 | 27.1±1.2 | 18.0 | 6.7±0.4 | 0.14±0.03 | 4U 1700+24 (107) | 256.644 | 23.972 | 0.1306 | LMXB/NS | BG | |
| 365 | 2MAXI J1708−810 | 257.2 | −81.0 | 0.312 | 7.3 | 6.0±0.8 | 4.5 | 1.2±0.3 | 0.26±0.12 | | | | | | | |
| 366 | 2MAXI J1712+641 | 258.2 | 64.1 | 0.076 | 8.8 | 8.9±1.1 | 11.2 | 3.4±0.3 | −0.08±0.07 | RXC J1712.7+6403 | 258.197 | 64.061 | 0.0037 | Galaxy Cluster | M | |
| 367 | 2MAXI J1715+201 | 258.9 | 20.1 | 0.213 | 7.0 | 6.6±0.9 | 4.7 | 1.7±0.4 | 0.14±0.13 | ZwCl8182 | 259.043 | 20.357 | | Sy2 | M | |
| 368 | 2MAXI J1717−628 | 259.3 | −62.9 | 0.075 | 14.8 | 14.2±1.0 | 1.3 | 0.4±0.3 | 0.85±0.10 | NGC 6300 (108) | 259.248 | −62.821 | 0.1644 | Galaxy Cluster | BG | |
| 369 | 2MAXI J1718−734 | 259.5 | −73.4 | 0.101 | 17.0 | 12.4±0.7 | 12.2 | 3.1±0.3 | 0.13±0.05 | | | | | | | |
| 370 | 2MAXI J1720+267 | 260.1 | 26.8 | 0.104 | 11.3 | 13.0±1.1 | 7.9 | 2.9±0.4 | 0.19±0.07 | RXC J1720.1+2637 | 260.039 | 26.627 | | Galaxy Cluster | M | |
| 371 | 2MAXI J1721+338 | 260.4 | 33.8 | 0.208 | 8.3 | 10.1±1.2 | 6.6 | 2.6±0.4 | 0.12±0.10 | | | | | | | |
| 372 | 2MAXI J1722+315 | 260.7 | 31.5 | 0.178 | 8.7 | 11.7±1.3 | 8.7 | 3.5±0.4 | 0.05±0.08 | | | | | | | |
| 373 | 2MAXI J1731−058 | 262.8 | −5.9 | 0.131 | 9.5 | 11.4±1.2 | 4.0 | 1.4±0.4 | 0.44±0.11 | 1RXS J173021.5−055933 | 262.591 | −5.992 | | CV/DQ Her | B | |
| 374 | 2MAXI J1740−599 | 265.1 | −60.0 | 0.154 | 10.7 | 9.9±0.9 | 8.2 | 2.3±0.3 | 0.16±0.07 | 1RXS J173751.2−600408 (C) () | 264.463 | −60.069 | | | B | (G) |
| 375 | 2MAXI J1741+059 | 265.3 | 5.9 | 0.115 | 10.5 | 12.8±1.2 | 9.8 | 3.6±0.4 | 0.07±0.07 | | | | | | | |
| 376 | 2MAXI J1741−539 | 265.4 | −54.0 | 0.156 | 7.4 | 7.4±1.0 | 5.1 | 1.6±0.3 | 0.21±0.11 | | | | | | | |
| 377 | 2MAXI J1742+037 | 265.5 | 3.8 | 0.101 | 12.2 | 13.5±1.1 | 11.2 | 4.1±0.4 | 0.03±0.06 | | | | | | | |
| 378 | 2MAXI J1742+184 | 265.6 | 18.5 | 0.098 | 11.4 | 13.6±1.2 | 10.2 | 3.8±0.4 | 0.08±0.07 | 4C +18.51 (109) | 265.529 | 18.456 | 0.186 | QSO | BG | |
| 379 | 2MAXI J1751−637 | 267.9 | −63.8 | 0.171 | 9.5 | 8.0±0.8 | 6.2 | 1.6±0.3 | 0.23±0.09 | RXC J1752.0−6348 | 268.023 | −63.816 | 0.133 | Galaxy Cluster | M | |
| 380 | 2MAXI J1753−013 | 268.4 | −1.3 | 0.004 | 255.1 | 495.9±1.9 | 192.3 | 145.3±0.8 | 0.06±0.00 | SWIFT J1753.5−0127 (110) | 268.368 | −1.452 | | LMXB/ BHC | BG | |
| 381 | 2MAXI J1759+684 | 269.9 | 68.5 | 0.206 | 9.3 | 6.8±0.7 | 5.9 | 1.5±0.3 | 0.20±0.11 | | | | | | | |
| 382 | 2MAXI J1800+082 | 270.1 | 8.2 | 0.077 | 13.9 | 15.7±1.1 | 6.2 | 2.3±0.4 | 0.38±0.08 | V2301 Oph | 270.146 | 8.170 | | CV/AM Her | B | |
| 383 | 2MAXI J1806+101 | 271.5 | 10.1 | 0.215 | 7.9 | 8.8±1.1 | 9.0 | 3.4±0.4 | −0.09±0.08 | RXC J1804.4+1002 | 271.119 | 10.039 | 0.1525 | Galaxy Cluster | M | |

TABLE 1 — Continued

| (1) | (2) | (3) | (4) | (5) | (6) | (7) | (8) | (9) | (10) | (11) | (12) | (13) | (14) | (15) | (16) | (17) |
|---|---|---|---|---|---|---|---|---|---|---|---|---|---|---|---|---|
| | MAXI | | | | | | | | | Counterpart | | | | | | |
| No. | Name | R.A. | Decl. | $\sigma_{stat}$[a] | $s_{D,4-10keV}$ | $f_{4-10keV}$[b] | $s_{D,3-4keV}$ | $f_{3-4keV}$[c] | HR | Name[d] | R.A. | Decl. | z | Type[e] | Flag[f] | Note[g] |
| 384 | 2MAXI J1807+059 | 271.9 | 6.0 | 0.051 | 22.5 | 25.9±1.2 | 13.6 | 5.3±0.4 | 0.24±0.04 | V426 Oph (111) | 271.966 | 5.864 | | CV / Dwarf nova | BG | |
| 385 | 2MAXI J1816+498 | 274.1 | 49.9 | 0.025 | 48.4 | 64.7±1.3 | 25.9 | 10.3±0.4 | 0.35±0.02 | AM Her (112) | 274.055 | 49.868 | | CV/AM Her | BG | |
| 386 | 2MAXI J1824+643 | 276.2 | 64.3 | 0.100 | 13.8 | 13.5±1.0 | 14.8 | 4.8±0.3 | -0.04±0.05 | | | | | | | |
| 387 | 2MAXI J1825+306 | 276.4 | 30.7 | 0.111 | 11.7 | 14.2±1.2 | 13.4 | 5.2±0.4 | -0.06±0.06 | RXC J1825.3+3026 (114) | 276.345 | 30.442 | 0.065 | Galaxy Cluster | MG | |
| 388 | 2MAXI J1825-370 | 276.5 | -37.0 | 0.006 | 219.1 | 399.4±1.8 | 106.5 | 63.5±0.6 | 0.35±0.00 | 4U 1822-371 (113) | 276.445 | -37.105 | | LMXB | BG | |
| 389 | 2MAXI J1834+327 | 278.7 | 32.8 | 0.059 | 21.8 | 29.1±1.3 | 16.7 | 7.3±0.4 | 0.13±0.04 | 3C 382 (115) | 278.764 | 32.696 | 0.0579 | Sy1 | BG | |
| 390 | 2MAXI J1835-328 | 279.0 | -32.9 | 0.029 | 41.7 | 55.7±1.3 | 34.9 | 15.3±0.4 | 0.09±0.02 | XB 1832-330 (116) | 278.932 | -32.991 | | LMXB/NS | BG | |
| 391 | 2MAXI J1836-594 | 279.0 | -59.4 | 0.070 | 15.6 | 14.8±0.9 | 13.9 | 4.0±0.3 | 0.09±0.05 | Fairall 49 | 279.243 | -59.402 | 0.0202 | Sy2 | B | |
| 392 | 2MAXI J1837-570 | 279.4 | -57.1 | 0.219 | 7.5 | 6.7±0.9 | 4.0 | 1.1±0.3 | 0.32±0.13 | SWIFT J183905.82-571507.6 | 279.776 | -57.252 | | SRC/X-RAY | B | |
| 393 | 2MAXI J1838-654 | 279.6 | -65.5 | 0.048 | 29.1 | 25.4±0.9 | 9.2 | 2.5±0.3 | 0.55±0.04 | ESO 103-035 (117) | 279.585 | -65.428 | 0.0133 | Sy2 | BG | |
| 394 | 2MAXI J1842+797 | 280.6 | 79.7 | 0.039 | 32.4 | 30.8±0.9 | 24.7 | 7.7±0.3 | 0.13±0.03 | 3C 390.3 (118) | 280.538 | 79.771 | 0.0561 | Sy1 | BMG | |
| 395 | 2MAXI J1844-373 | 281.1 | -37.3 | 0.196 | 7.4 | 8.3±1.1 | 2.8 | 1.0±0.4 | 0.45±0.15 | RXC J1844.5-3724 | 281.132 | -37.405 | 0.203 | Galaxy Cluster | M | |
| 396 | 2MAXI J1845-626 | 281.4 | -62.6 | 0.158 | 8.5 | 7.1±0.8 | 5.3 | 1.5±0.3 | 0.21±0.11 | Fairall 51 | 281.225 | -62.365 | 0.0142 | Sy1 | B | |
| 397 | 2MAXI J1852-783 | 283.0 | -78.4 | 0.065 | 19.7 | 16.3±0.8 | 15.1 | 3.9±0.3 | 0.15±0.04 | | | | | | | |
| 398 | 2MAXI J1854-310 | 283.7 | -31.1 | 0.021 | 58.1 | 77.6±1.3 | 33.0 | 13.1±0.4 | 0.32±0.02 | V1223 Sgr (120) | 283.759 | -31.163 | | CV/DQ Her | BG | |
| 399 | 2MAXI J1858+290 | 284.6 | 29.1 | 0.152 | 8.5 | 10.3±1.2 | 7.4 | 2.8±0.4 | 0.09±0.09 | | | | | | | |
| 400 | 2MAXI J1900-248 | 285.1 | -24.8 | 0.004 | 203.9 | 322.0±1.6 | 149.7 | 95.3±0.6 | 0.05±0.00 | HETE J1900.1-2455 (121) | 285.036 | -24.921 | | LMXB/msPSR | BG | |
| 401 | 2MAXI J1900-748 | 285.1 | -74.8 | 0.256 | 8.7 | 6.6±0.8 | 8.6 | 2.1±0.2 | 0.01±0.08 | | | | | | | |
| 402 | 2MAXI J1912-194 | 288.1 | -19.5 | 0.195 | 7.9 | 8.5±1.1 | 9.1 | 3.1±0.3 | -0.05±0.08 | | | | | | | |
| 403 | 2MAXI J1915-626 | 288.8 | -62.6 | 0.199 | 9.7 | 7.8±0.8 | 6.4 | 1.7±0.3 | 0.19±0.09 | | | | | | | |
| 404 | 2MAXI J1919-525 | 289.9 | -52.6 | 0.177 | 9.0 | 9.0±1.0 | 8.2 | 2.6±0.3 | 0.05±0.08 | | | | | | | |
| 405 | 2MAXI J1919-297 | 290.0 | -29.8 | 0.192 | 7.8 | 8.6±1.1 | 4.2 | 1.5±0.4 | 0.31±0.12 | PKS 1916-300 | 289.867 | -29.969 | 0.1668 | Sy1.9 | B | |
| 406 | 2MAXI J1921-587 | 290.3 | -58.8 | 0.046 | 27.0 | 26.2±1.0 | 25.7 | 8.0±0.3 | 0.04±0.03 | ESO 141-G 055 (122) | 290.309 | -58.670 | 0.036 | Sy1 | BG | |
| 407 | 2MAXI J1921+440 | 290.3 | 44.0 | 0.018 | 68.8 | 108.6±1.6 | 54.4 | 30.3±0.6 | 0.08±0.01 | ABELL 2319 (123) | 290.189 | 43.962 | 0.0557 | Galaxy Cluster | BMG | |
| 408 | 2MAXI J1922-171 | 290.6 | -17.1 | 0.034 | 36.8 | 42.9±1.2 | 37.4 | 14.3±0.4 | -0.01±0.02 | SWIFT J1922.7-1716 | 290.654 | -17.284 | | XRB/ uQuasar? | B | |
| 409 | 2MAXI J1924+502 | 291.1 | 50.3 | 0.124 | 9.1 | 12.2±1.3 | 4.4 | 1.7±0.4 | 0.41±0.10 | CH Cyg (124) | 291.138 | 50.241 | | Symb/WD | BG | |
| 410 | 2MAXI J1929-497 | 292.4 | -49.8 | 0.156 | 8.6 | 9.9±1.2 | 7.3 | 2.1±0.3 | 0.21±0.09 | | | | | | | |
| 411 | 2MAXI J1933-802 | 293.3 | -80.2 | 0.093 | 13.6 | 11.2±0.8 | 13.6 | 3.6±0.3 | 0.01±0.05 | | | | | | | |
| 412 | 2MAXI J1936-513 | 294.1 | -51.3 | 0.124 | 10.5 | 11.3±1.1 | 8.2 | 2.7±0.3 | 0.15±0.08 | 2MASX J19380437-5109497 | 294.518 | -51.164 | 0.04 | Sy1.2 | B | |
| 413 | 2MAXI J1937-060 | 294.3 | -6.0 | 0.075 | 16.7 | 19.1±1.1 | 15.6 | 5.8±0.4 | 0.04±0.04 | 2MASX J19373299-0613046 | 294.388 | -6.218 | 0.0103 | Sy1.5 | B | |
| 414 | 2MAXI J1941-102 | 295.4 | -10.3 | 0.034 | 37.4 | 44.6±1.2 | 22.3 | 8.1±0.4 | 0.29±0.02 | | | | | | | |
| 415 | 2MAXI J1947-766 | 296.9 | -76.7 | 0.116 | 10.6 | 8.7±0.8 | 8.1 | 2.0±0.3 | 0.17±0.08 | RXC J1947.3-7623 | 296.830 | -76.392 | 0.217 | Galaxy Cluster | M | |
| 416 | 2MAXI J1950-522 | 297.7 | -52.2 | 0.144 | 9.7 | 10.1±1.0 | 8.4 | 2.8±0.3 | 0.09±0.08 | | | | | | | |
| 417 | 2MAXI J1950-110 | 297.7 | -11.0 | 0.177 | 7.2 | 8.3±1.1 | 6.2 | 2.1±0.3 | 0.13±0.10 | | | | | | | |
| 418 | 2MAXI J1952+708 | 298.1 | 70.8 | 0.160 | 9.1 | 7.9±0.9 | 5.0 | 1.4±0.3 | 0.28±0.11 | | | | | | | |
| 419 | 2MAXI J1954-554 | 298.5 | -55.4 | 0.248 | 8.0 | 6.6±0.8 | 10.3 | 3.3±0.3 | -0.21±0.08 | RXC J1952.2-5503 | 298.069 | -55.062 | 0.06 | Galaxy Cluster | M | |
| 420 | 2MAXI J1959+651 | 299.9 | 65.1 | 0.031 | 40.6 | 42.9±1.1 | 48.6 | 18.2±0.4 | -0.13±0.02 | QSO B1959+650 (126) | 299.999 | 65.148 | 0.047 | BL Lac | BG | |
| 421 | 2MAXI J2005-712 | 301.4 | -71.2 | 0.129 | 9.7 | 7.4±0.8 | 7.0 | 1.7±0.2 | 0.17±0.09 | | | | | | | |
| 422 | 2MAXI J2006-107 | 301.6 | -10.7 | 0.198 | 7.4 | 7.7±1.0 | 4.6 | 1.5±0.3 | 0.24±0.12 | | | | | | | |
| 423 | 2MAXI J2009-487 | 302.3 | -48.8 | 0.114 | 9.9 | 11.1±1.1 | 11.1 | 4.0±0.4 | -0.05±0.07 | PKS 2005-489 | 302.356 | -48.832 | 0.071 | BL Lac | B | |
| 424 | 2MAXI J2009-611 | 302.4 | -61.1 | 0.089 | 13.2 | 12.1±0.9 | 12.1 | 3.4±0.3 | 0.08±0.06 | NGC 6860 (127) | 302.195 | -61.100 | 0.0149 | Sy1 | BG | |
| 425 | 2MAXI J2011+604 | 303.0 | 60.4 | 0.205 | 7.9 | 8.3±1.0 | 7.8 | 2.5±0.3 | 0.03±0.09 | | | | | | | |
| 426 | 2MAXI J2012-568 | 303.2 | -56.8 | 0.024 | 52.3 | 51.5±1.0 | 45.0 | 15.4±0.3 | 0.04±0.01 | ABELL 3667 (128) | 303.125 | -56.816 | 0.0556 | Galaxy Cluster | BMG | |
| 427 | 2MAXI J2014-244 | 303.6 | -24.4 | 0.136 | 9.0 | 10.0±1.1 | 7.7 | 2.6±0.3 | 0.10±0.08 | RXC J2014.8-2430 | 303.707 | -24.508 | 0.1612 | Galaxy Cluster | M | |
| 428 | 2MAXI J2018+156 | 304.7 | 15.7 | 0.278 | 7.0 | 9.4±1.3 | 7.1 | 2.7±0.4 | 0.06±0.10 | | | | | | | |
| 429 | 2MAXI J2033+219 | 308.4 | 22.0 | 0.132 | 10.3 | 12.1±1.2 | 6.5 | 2.3±0.4 | 0.26±0.09 | 4C +21.55 | 308.384 | 21.773 | 0.1735 | QSO | B | |
| 430 | 2MAXI J2034-353 | 308.6 | -35.4 | 0.132 | 12.0 | 14.3±1.2 | 12.9 | 4.7±0.4 | -0.01±0.06 | | | | | | | |
| 431 | 2MAXI J2037-306 | 309.4 | -30.7 | 0.142 | 10.1 | 11.2±1.1 | 4.1 | 1.4±0.3 | 0.44±0.11 | | | | | | | |

TABLE 1 — Continued

| (1) | (2) | (3) | (4) | (5) | (6) | (7) | (8) | (9) | (10) | (11) | (12) | (13) | (14) | (15) | (16) | (17) |
|---|---|---|---|---|---|---|---|---|---|---|---|---|---|---|---|---|
| | MAXI | | | | | | | | | Counterpart | | | | | | |
| No. | Name | R.A. | Decl. | $\sigma_{\rm stat}$[a] | $s_{\rm D,4-10keV}$ | $f_{\rm 4-10keV}$[b] | $s_{\rm D,3-4keV}$ | $f_{\rm 3-4keV}$[c] | HR | Name[d] | R.A. | Decl. | z | Type[e] | Flag[f] | Note[g] |
| 432 | 2MAXI J2042+751 | 310.7 | 75.2 | 0.055 | 22.9 | 21.4±0.9 | 17.4 | 5.2±0.3 | 0.15±0.04 | 4C +74.26 (129) | 310.655 | 75.134 | 0.104 | Sy1 | BG | |
| 433 | 2MAXI J2043-324 | 310.8 | -32.5 | 0.194 | 7.2 | 7.2±1.0 | 10.2 | 3.7±0.4 | -0.22±0.08 | | | | | | | |
| 434 | 2MAXI J2043-714 | 310.9 | -71.4 | 0.212 | 8.0 | 6.9±0.9 | 8.0 | 2.0±0.3 | 0.06±0.09 | | | | | | BG | |
| 435 | 2MAXI J2044-106 | 311.1 | -10.6 | 0.035 | 36.1 | 42.5±1.2 | 31.4 | 11.6±0.4 | 0.09±0.02 | Mrk 509 (130) | 311.041 | -10.723 | 0.0344 | Sy1.2 | | |
| 436 | 2MAXI J2107-217 | 316.8 | -21.7 | 0.123 | 8.9 | 7.7±0.9 | 6.3 | 2.0±0.3 | 0.11±0.10 | | | | | | | |
| 437 | 2MAXI J2108-439 | 317.1 | -43.9 | 0.230 | 7.8 | 9.5±1.2 | 3.2 | 1.2±0.4 | 0.44±0.14 | | | | | | | |
| 438 | 2MAXI J2114-588 | 318.6 | -58.9 | 0.130 | 9.3 | 8.2±0.9 | 7.5 | 2.1±0.3 | 0.13±0.08 | CD Ind | 318.904 | -58.698 | | CV/AM HER | B | |
| 439 | 2MAXI J2115+821 | 318.8 | 82.1 | 0.137 | 10.1 | 9.6±0.9 | 9.3 | 2.7±0.3 | 0.07±0.07 | 2MASX J21140128+8204483 | 318.505 | 82.080 | 0.084 | Sy1 | B | |
| 440 | 2MAXI J2130+122 | 322.5 | 12.2 | 0.011 | 105.7 | 166.9±1.6 | 97.0 | 54.0±0.6 | 0.01±0.01 | 4U 2129+12 (131) | 322.493 | 12.167 | | LMXB in globular clus | BG | |
| 441 | 2MAXI J2131-334 | 322.9 | -33.5 | 0.220 | 7.3 | 8.4±1.1 | 6.7 | 2.4±0.4 | 0.07±0.10 | 6dF J2132022-334254 | 323.009 | -33.715 | 0.0293 | Sy1 | B | |
| 442 | 2MAXI J2136-624 | 324.1 | -62.4 | 0.061 | 21.1 | 18.4±0.9 | 14.6 | 4.1±0.3 | 0.20±0.04 | 1RXS J213623.1-622400 (132) | 324.096 | -62.400 | 0.0588 | Sy1 | BG | |
| 443 | 2MAXI J2138-643 | 324.5 | -64.3 | 0.122 | 10.9 | 9.6±0.9 | 9.6 | 2.5±0.3 | 0.10±0.07 | | | | | | | |
| 444 | 2MAXI J2144+383 | 326.2 | 38.4 | <0.001 | 1171.9 | 6549.6±5.6 | 781.7 | 2176.6±2.8 | -0.01±0.00 | Cyg X-2 (133) | 326.172 | 38.322 | | LMXB/NS | BG | |
| 445 | 2MAXI J2147-693 | 326.8 | -69.4 | 0.231 | 8.2 | 5.6±0.7 | 6.0 | 1.5±0.3 | 0.10±0.10 | | | | | | | |
| 446 | 2MAXI J2148-073 | 327.1 | -7.4 | 0.191 | 8.1 | 8.8±1.1 | 5.3 | 1.9±0.4 | 0.21±0.11 | | | | | | | |
| 447 | 2MAXI J2148-166 | 327.2 | -16.7 | 0.205 | 7.4 | 4.9±0.7 | 4.9 | 1.5±0.3 | 0.02±0.12 | | | | | | | |
| 448 | 2MAXI J2150+142 | 327.5 | 14.2 | 0.171 | 10.2 | 12.4±1.2 | 7.2 | 2.8±0.4 | 0.18±0.08 | | | | | | | |
| 449 | 2MAXI J2150-192 | 327.7 | -19.2 | 0.113 | 11.7 | 12.1±1.0 | 11.4 | 3.6±0.3 | 0.04±0.06 | 6dF J2149581-185924 | 327.492 | -18.990 | 0.1581 | Sy1 | B | |
| 450 | 2MAXI J2151-723 | 327.8 | -72.3 | 0.196 | 7.8 | 5.4±0.7 | 5.1 | 1.3±0.2 | 0.17±0.11 | | | | | | | |
| 451 | 2MAXI J2151-304 | 328.0 | -30.5 | 0.099 | 12.7 | 15.3±1.2 | 10.8 | 3.9±0.4 | 0.12±0.06 | PKS 2149-306 | 327.981 | -30.465 | 2.345 | Blazar | B | |
| 452 | 2MAXI J2152-576 | 328.1 | -57.7 | 0.088 | 14.7 | 13.4±0.9 | 13.8 | 4.3±0.3 | 0.01±0.05 | | | | | | | |
| 453 | 2MAXI J2154+178 | 328.7 | 17.8 | 0.108 | 12.1 | 14.7±1.2 | 11.2 | 4.3±0.4 | 0.06±0.06 | RXC J2153.5+1741 | 328.398 | 17.687 | 0.2329 | Galaxy Cluster | M | |
| 454 | 2MAXI J2157-061 | 329.5 | -6.1 | 0.181 | 7.2 | 8.2±1.1 | 6.0 | 2.1±0.4 | 0.11±0.11 | EUVE J2157-06.1 | 329.363 | -6.173 | | Star | B | |
| 455 | 2MAXI J2200-600 | 330.2 | -60.0 | 0.059 | 20.7 | 19.4±0.9 | 19.5 | 5.7±0.3 | 0.06±0.04 | RXC J2201.9-5956 (134) | 330.483 | -59.949 | 0.098 | Galaxy Cluster | MG | |
| 456 | 2MAXI J2200+105 | 330.2 | 10.6 | 0.153 | 8.2 | 9.6±1.2 | 3.7 | 1.3±0.4 | 0.41±0.12 | Mrk 520 | 330.172 | 10.552 | 0.0266 | Sy1.9 | B | |
| 457 | 2MAXI J2201-317 | 330.5 | -31.8 | 0.065 | 20.8 | 25.2±1.2 | 6.4 | 2.2±0.4 | 0.57±0.05 | NGC 7172 (135) | 330.508 | -31.870 | 0.0087 | Sy2 | BG | |
| 458 | 2MAXI J2209-125 | 332.4 | -12.5 | 0.107 | 11.9 | 13.4±1.1 | 12.4 | 4.2±0.3 | 0.02±0.06 | | | | | | | |
| 459 | 2MAXI J2213-891 | 333.3 | -89.2 | 0.264 | 7.1 | 5.7±0.8 | 5.5 | 1.5±0.3 | 0.11±0.11 | | | | | | | |
| 460 | 2MAXI J2216-032 | 334.2 | -3.3 | 0.131 | 11.0 | 12.5±1.1 | 10.6 | 3.9±0.4 | 0.03±0.07 | | | | | | | |
| 461 | 2MAXI J2216-647 | 334.2 | -64.8 | 0.223 | 7.6 | 5.5±0.7 | 5.3 | 1.4±0.3 | 0.13±0.11 | RXC J2218.0-6511 | 334.523 | -65.185 | 0.0951 | Galaxy Cluster | M | |
| 462 | 2MAXI J2218-083 | 334.5 | -8.4 | 0.047 | 28.3 | 33.6±1.2 | 10.4 | 3.7±0.4 | 0.49±0.04 | FO AQR (137) | 334.481 | -8.351 | | CV/DQ Her | BG | |
| 463 | 2MAXI J2218-388 | 334.7 | -38.9 | 0.143 | 9.7 | 9.8±1.0 | 7.5 | 2.7±0.4 | 0.08±0.08 | RXC J2218.6-3853 | 334.668 | -38.898 | 0.1411 | Galaxy Cluster | M | |
| 464 | 2MAXI J2223-017 | 336.0 | -1.8 | 0.089 | 12.7 | 14.7±1.2 | 8.9 | 3.2±0.4 | 0.20±0.07 | RXC J2223.8-0138 | 335.970 | -1.638 | 0.0906 | Galaxy Cluster | M | |
| 465 | 2MAXI J2225+012 | 336.5 | 1.2 | 0.194 | 7.1 | 8.4±1.2 | 4.2 | 1.5±0.4 | 0.29±0.13 | | | | | | | |
| 466 | 2MAXI J2235-259 | 338.8 | -25.9 | 0.065 | 19.3 | 22.5±1.2 | 15.8 | 5.6±0.4 | 0.14±0.04 | | | | | | | |
| 467 | 2MAXI J2235-381 | 339.0 | -38.1 | 0.152 | 13.9 | 14.0±1.0 | 11.4 | 4.3±0.4 | 0.03±0.06 | | | | | | | |
| 468 | 2MAXI J2238-126 | 339.6 | -12.6 | 0.167 | 8.1 | 8.4±1.0 | 6.0 | 2.0±0.3 | 0.17±0.10 | Mrk 915 | 339.194 | -12.545 | 0.0241 | Sy1 | B | |
| 469 | 2MAXI J2241+297 | 340.5 | 29.7 | 0.148 | 7.5 | 8.4±1.1 | 12.0 | 4.5±0.4 | -0.24±0.07 | | | | | | | |
| 470 | 2MAXI J2247-524 | 341.9 | -52.5 | 0.096 | 13.3 | 14.0±1.1 | 12.0 | 4.0±0.3 | 0.07±0.06 | | | | | | | |
| 471 | 2MAXI J2249-641 | 342.4 | -64.2 | 0.131 | 8.7 | 7.7±0.9 | 11.3 | 3.0±0.3 | -0.08±0.07 | RXC J2249.9-6425 | 342.488 | -64.429 | 0.094 | Galaxy Cluster | M | |
| 472 | 2MAXI J2250-445 | 342.5 | -44.6 | 0.163 | 8.9 | 11.9±1.3 | 8.7 | 3.4±0.4 | 0.06±0.08 | RXC J2248.7-4431 | 342.181 | -44.529 | 0.3475 | Galaxy Cluster | M | |
| 473 | 2MAXI J2252-323 | 343.2 | -32.4 | 0.123 | 9.5 | 10.5±1.1 | 8.2 | 2.9±0.4 | 0.08±0.08 | | | | | | | |
| 474 | 2MAXI J2253+166 | 343.4 | 16.7 | 0.049 | 22.9 | 30.6±1.3 | 24.5 | 10.7±0.4 | -0.03±0.03 | | | | | | | |
| 475 | 2MAXI J2254-174 | 343.5 | -17.4 | 0.039 | 28.9 | 33.0±1.1 | 25.2 | 8.9±0.4 | 0.10±0.03 | MR 2251-178 (140) | 343.524 | -17.582 | 0.064 | Sy1 | BG | |
| 476 | 2MAXI J2255-030 | 343.8 | -3.0 | 0.042 | 31.4 | 38.1±1.2 | 17.1 | 6.5±0.4 | 0.32±0.03 | | | | | | | |
| 477 | 2MAXI J2255-274 | 343.9 | -27.4 | 0.216 | 7.2 | 8.1±1.1 | 4.6 | 1.6±0.3 | 0.24±0.12 | | | | | | | |
| 478 | 2MAXI J2300+250 | 345.0 | 25.0 | 0.156 | 9.7 | 10.1±1.0 | 6.7 | 2.4±0.4 | 0.16±0.09 | KAZ 320 | 344.887 | 24.918 | 0.0345 | Sy1 | B | |
| 479 | 2MAXI J2301-591 | 345.3 | -59.1 | 0.177 | 8.3 | 6.8±0.8 | 4.5 | 1.2±0.3 | 0.29±0.12 | 2MASX J23013626-5913210 | 345.401 | -59.222 | 0.149 | Sy1 | B | |

TABLE 1 — CONTINUED

| (1) | (2) | (3) | (4) | (5) | (6) | (7) | (8) | (9) | (10) | (11) | (12) | (13) | (14) | (15) | (16) | (17) |
|---|---|---|---|---|---|---|---|---|---|---|---|---|---|---|---|---|
| | MAXI | | | | | | | | | Counterpart | | | | | | |
| No. | Name | R.A. | Decl. | $\sigma_{stat}$[a] | $s_{D,4-10keV}$ | $f_{4-10keV}$[b] | $s_{D,3-4keV}$ | $f_{3-4keV}$[c] | HR | Name[d] | R.A. | Decl. | z | Type[e] | Flag[f] | Note[g] |
| 480 | 2MAXI J2303+086 | 345.8 | 8.7 | 0.071 | 17.8 | 21.7±1.2 | 16.9 | 6.6±0.4 | 0.03±0.04 | NGC 7469 | 345.815 | 8.874 | 0.0163 | Sy1.2 | BM | |
| 481 | 2MAXI J2304−187 | 346.1 | −18.8 | 0.177 | 7.0 | 7.4±1.1 | 4.5 | 1.5±0.3 | 0.25±0.12 | PKS 2300−18 | 345.762 | −18.690 | 0.1283 | Sy1 | B | |
| 482 | 2MAXI J2304−085 | 346.2 | −8.6 | 0.040 | 31.3 | 38.0±1.2 | 27.1 | 10.5±0.4 | 0.09±0.02 | Mrk 926 (142) | 346.181 | −8.686 | 0.0469 | Sy1.5 | BG | |
| 483 | 2MAXI J2312−220 | 348.0 | −22.1 | 0.131 | 9.6 | 10.3±1.1 | 9.0 | 3.0±0.3 | 0.05±0.08 | | | | | | | |
| 484 | 2MAXI J2316−424 | 349.1 | −42.4 | 0.124 | 10.5 | 14.1±1.3 | 9.0 | 3.5±0.4 | 0.13±0.07 | | | | | | | |
| 485 | 2MAXI J2318+001 | 349.5 | 0.2 | 0.120 | 11.1 | 13.4±1.2 | 9.6 | 3.6±0.4 | 0.10±0.07 | NGC 7603 | 349.736 | 0.244 | 0.0295 | Sy1.5 | B | |
| 486 | 2MAXI J2318+188 | 349.7 | 18.8 | 0.150 | 8.5 | 10.4±1.2 | 12.7 | 4.6±0.4 | −0.15±0.07 | RXC J2318.5+1842 (C) | 349.612 | 18.724 | 0.0389 | Galaxy Cluster | M | |
| 487 | 2MAXI J2319+421 | 349.9 | 42.1 | 0.142 | 9.4 | 12.5±1.3 | 8.9 | 3.9±0.4 | 0.03±0.08 | RXC J2320.2+4146 | 350.060 | 41.779 | 0.14 | Galaxy Cluster | M | (H) |
| 488 | 2MAXI J2323+221 | 350.8 | 22.1 | 0.245 | 7.5 | 7.8±1.0 | 1.9 | 0.7±0.3 | 0.58±0.18 | | | | | | | |
| 489 | 2MAXI J2324+166 | 351.0 | 16.7 | 0.153 | 10.3 | 12.4±1.2 | 9.8 | 3.7±0.4 | 0.05±0.07 | A2589 | 350.973 | 16.809 | 0.0416 | Galaxy Cluster | M | |
| 490 | 2MAXI J2324−121 | 351.1 | −12.1 | 0.111 | 11.4 | 11.3±1.0 | 11.7 | 4.1±0.4 | −0.05±0.06 | RXC J2325.3−1207 | 351.333 | −12.127 | 0.0852 | Galaxy Cluster | M | |
| 491 | 2MAXI J2330−022 | 352.7 | −2.2 | 0.192 | 7.8 | 9.1±1.2 | 6.2 | 2.2±0.4 | 0.15±0.10 | | | | | | | |
| 492 | 2MAXI J2339+269 | 354.8 | 27.0 | 0.178 | 10.2 | 10.3±1.0 | 12.4 | 4.6±0.4 | −0.16±0.06 | RXC J2338.4+2659 | 354.607 | 27.012 | 0.0309 | Galaxy Cluster | M | |
| 493 | 2MAXI J2344+092 | 356.2 | 9.3 | 0.110 | 10.3 | 12.5±1.2 | 12.7 | 4.9±0.4 | −0.09±0.06 | RXC J2344.9+0911 | 356.238 | 9.198 | 0.04 | Galaxy Cluster | M | |
| 494 | 2MAXI J2348−280 | 357.1 | −28.0 | 0.079 | 15.9 | 18.9±1.2 | 20.0 | 7.5±0.4 | −0.09±0.04 | RXC J2347.7−2808 | 356.930 | −28.141 | 0.03 | Galaxy Cluster | M | |
| 495 | 2MAXI J2354+286 | 358.6 | 28.6 | 0.119 | 10.3 | 12.4±1.2 | 13.8 | 5.3±0.4 | −0.13±0.06 | | | | | | | |
| 496 | 2MAXI J2354−169 | 358.7 | −16.9 | 0.156 | 9.1 | 8.9±1.0 | 6.1 | 2.0±0.3 | 0.18±0.10 | | | | | | | |
| 497 | 2MAXI J2356−106 | 359.2 | −10.7 | 0.252 | 7.4 | 7.4±1.0 | 6.6 | 2.3±0.3 | 0.03±0.10 | | | | | | | |
| 498 | 2MAXI J2358−346 | 359.6 | −34.7 | 0.105 | 13.4 | 15.7±1.2 | 16.5 | 6.3±0.4 | −0.10±0.05 | | | | | | | |
| 499 | 2MAXI J2359−307 | 359.8 | −30.8 | 0.121 | 10.2 | 11.7±1.1 | 9.9 | 3.5±0.4 | 0.04±0.07 | H 2356−309 | 359.783 | −30.628 | 0.1651 | BL Lac | B | |
| 500 | 2MAXI J2359−068 | 359.9 | −6.8 | 0.195 | 7.1 | 7.5±1.1 | 4.2 | 1.5±0.4 | 0.23±0.13 | 2MASX J00004876−0709117 | 0.203 | −7.153 | | Galaxy | B | |

[a] The $1\sigma$ statistical position error in units of degree. There exists an additional systematic error of $\simeq 0.09°$ (90%). See section 4.4 for details.
[b] The observed 4–10 keV flux in units of $10^{-12}$ ergs cm$^{-2}$ s$^{-1}$, converted from Crab units by assuming the Crab-like spectrum; 1 mCrab = $1.21 \times 10^{-11}$ ergs cm$^{-2}$ s$^{-1}$.
[c] The observed 3–4 keV flux in units of $10^{-12}$ ergs cm$^{-2}$ s$^{-1}$, converted from Crab units by assuming the Crab-like spectrum; 1 mCrab = $3.98 \times 10^{-12}$ ergs cm$^{-2}$ s$^{-1}$.
[d] The numbers in parentheses represent source identification ones in the GSC7 catalog.
[e] Sy: Seyfert galaxy, L(H)MXB: low(high)-mass X-ray binary, Symb: symbiotic star, WD: white dwarf, CV: cataclysmic variable, NS: neutron star, BH: black hole candidate, msPSR: millisecond pulsar, FSRQ: flat-spectrum radio quasar.
[f] Cross-matching flag; B, M, and G represent that the source has more one or more counterparts in the BAT70, MCXC, and GSC7 catalog, respectively. The counterpart name listed in column (11) is quoted from the catalog expressed by the leftmost letter in this column.
[g] (A) We change the counterpart name from SWIFT J0042.7+4111 to M31 taking into account the angular resolution of MAXI/GSC. (B) RXC J0105.5−2439 is another counterpart of this source. (C) We change the counterpart name from NGC 1275 to Perseus Cluster taking into account the angular resolution of MAXI/GSC. (D) We change the counterpart name from CXO J051406.4−400238 to NGC 1851 taking into account the angular resolution of MAXI/GSC. (E) Since MAXI J0957+693 is a confused source consisting of M81, M82, and 1RXS J095755.0+690310, we use the position determined by the MAXI/GSC 7-month survey for the cross-matching. (F) Since the position accuracy of 3EG J1627−2419, the counterpart of MAXI J0957+693, is low, we use the position determined by the MAXI/GSC 7-month survey for the cross-matching. (G) ESO 139−G 012 is another counterpart of this source. (H) ABELL 2572b is another counterpart of this source.